\documentclass[aps,prl,twocolumn,noshowpacs,superscriptaddress]{revtex4-1}
\usepackage{graphicx,epsfig, subfigure}
\usepackage{amsbsy}
\usepackage{amssymb}
\usepackage{amsmath}
\usepackage{hyperref}
\usepackage[bottom]{footmisc}
\usepackage{xcolor}
\usepackage{color}
\usepackage{times}
\usepackage{textcomp}
\usepackage{verbatim} 

\date{\today}

\begin{document}
\newcommand{\eqnref}[1]{Eq.~\ref{#1}}
\newcommand{\figref}[2][]{Fig.~\ref{#2}#1}
\newcommand{\RN}[1]{%
  \textup{\uppercase\expandafter{\romannumeral#1}}%
}

\title{Quantum information processing with closely-spaced diamond color centers \\
in strain and magnetic fields}

\author{Zhujing Xu}
	\affiliation{Department of Physics and Astronomy, Purdue University, West Lafayette, Indiana 47907, USA}
	\author{Zhang-qi Yin}
	\affiliation{Center for Quantum Technology Research, School of Physics, Beijing Institute of Technology, Beijing 100081, China} 
	\author{Qinkai Han}
	\affiliation{State Key Laboratory of Tribology, Deparment of Mechanical Engineering, Tsinghua University, Beijing 100084, China}
	\author{Tongcang Li}
	\email{tcli@purdue.edu}
	\affiliation{Department of Physics and Astronomy, Purdue University, West Lafayette, Indiana 47907, USA}
	\affiliation{School of Electrical and Computer Engineering, Purdue University, West Lafayette, Indiana 47907, USA}
	\affiliation{Birck Nanotechnology Center, Purdue University, West Lafayette, Indiana 47907, USA}
	\affiliation{Purdue Quantum Science and Engineering Institute, Purdue University, West Lafayette, Indiana 47907, USA}
	\date{\today}


\begin{abstract}
Electron and nuclear spins of diamond nitrogen-vacancy (NV) centers are good candidates for quantum information processing as they have long coherence time and can be initialized and read out optically. However, creating a large number of coherently coupled and individually addressable NV centers for quantum computing has been a big challenge. Here we propose methods to use high-density diamond NV centers coupled by spin-spin interaction with an average separation on the order of $10$~nm for quantum computing. We propose to use a strain gradient to encode the position information of each NV center in the energy level of its excited electron orbital state, which causes a shift of its optical transition frequency. With such strain encoding, more than 100 closely-packed NV centers below optical diffraction limit can be read out individually by resonant optical excitation. A magnetic gradient will be used to shift the electron spin resonant (ESR) frequencies of  NV centers. Therefore, the spin state of each NV center can be individually manipulated and different NV centers can be selectively coupled. A universal set of quantum operations for two-qubit and three-qubit system is introduced by careful design of external drives.
Moreover, entangled states with multiple qubits can be created by this protocol, which is a major step towards quantum information processing with solid-state spins.
\end{abstract}

\maketitle

\section{Introduction}
A scalable quantum computer and network requires the long coherence time of single qubit, capability of initialization and full control, as well as the scalability and correctibility\cite{Matthews:2009aa,P.:2000aa}. Superconducting circuits\cite{Song:2017aa}, trapped ions\cite{Schafer:2018aa}, quantum dots\cite{Press:2008aa, Delteil:2017aa} and integrated photonic devices\cite{Caspani:2017aa} are some promising candidates for quantum computing. Electron and nuclear spins of diamond nitrogen-vacancy (NV) centers are also good platforms for quantum computing, for the remarkable fact that they have long coherence time\cite{Bar-Gill:2013aa,Lange:2010aa} and can be initialized with high fidelity and read out optically    \cite{Neumann:2010aa,Jelezko:2004aa,Gruber:1997aa}.  Besides, the solid-state material has the advantage for developing integrated device system.
Several quantum operations have already been demonstrated at even room temperature for both electron spins\cite{Hanson:2006aa} and nuclear spins\cite{Jelezko:2004aa} in diamond, which enabled the long-range entanglement\cite{Hensen:2015aa, Humphreys2017}, nanoscale sensing\cite{Mamin:2013aa}, quantum information\cite{Maurer:2012aa, Yao:2011aa} and quantum simulations\cite{Cai:2013aa, Wang:2015aa}. Recently, robust universal quantum gates with high fidelity for single-qubit system is accomplished by a novelly designed control pulse sequence\cite{Rong:2015aa}. Entanglement between two NV electron spins has been realized\cite{Dolde:2013aa} and improved to achieve high fidelity by utilizing nuclear spins\cite{Dolde:2014aa}, which provides the promising scalability of NV centers.  Various designs and protocols for scalable structure of NV centers are also carried out\cite{NYao2012scalable, Greentree:2016aa, Schroder:2017aa}. Progress towards an integrated system with fault-tolerant multi-qubit gates is unceasing. However, creating a large number of  coherently coupled and individually addressable NV centers has been a big challenge.

\begin{figure*}[t]
	\begin{center}
	\includegraphics[scale=0.45]{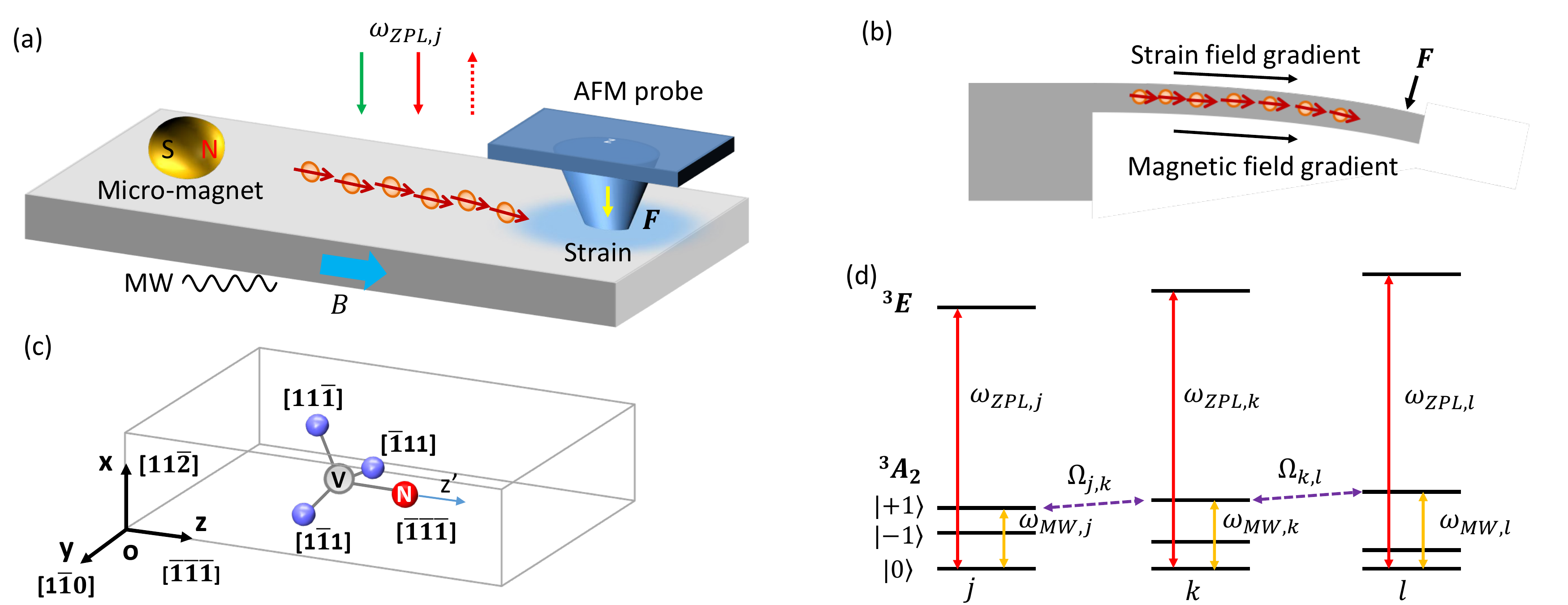}
	\caption{Schematic of quantum computing with high-density diamond NV centers via strain and magnetic encoding. (a). By the contact stress from an AFM tip, an inhomogeneous strain field is applied on the diamond cantilever. Such a gradient strain field can shift the optical transition frequencies of the NV centers depending on their positions. An external gradient magnetic field is applied on the diamond to split the ODMR frequencies of each NV center. Well-designed global composite microwave (MW) pulses with multiple frequencies are applied on the diamond to control the electron spins individually. (b). The gradient strain field and the magnetic field are both along the long axis of the cantilever. (c). A NV center is embedded in a diamond cantilever. The red sphere denotes the nitrogen atom and the blue are for the carbon atoms. The gradient strain field and the magnetic field are applied along z axis in the cantilever coordinates, which is [$\bar1\bar1\bar1$] in Miller indices for diamonds. $z'$ denotes direction of the NV axis. One possible orientation of the NV axis is [$\bar1\bar1\bar1$] as shown in the figure. The other three are [$11\bar1$], [$1\bar11$] and [$\bar111$].
	(d). Energy splits of optical transition frequencies and ODMR frequencies, due to strain field interaction and Zeeman effect, respectively. The frequencies of each NV center are designed to be unmatched with any other spins. Individual control and readout can be realized by this approach.
	}
	\label{scheme}
	\end{center}
\end{figure*}

Here we consider an approach to individually manipulate and readout, and coherently couple more than 100 closely packed diamond NV centers. An inhomogeneous strain field can be applied by an atomic force microscope (AFM) probe onto the diamond surface as shown in Fig.\ref{scheme}.(a).  An alternative way to realize the strain field gradient is to apply an external static force at one end of a diamond cantilever while fix the other end (Fig.\ref{scheme}.(b)).  The strain field generated by this method will be more controllable than the randomly created inhomogeneous strain field in a polycrystalline diamond \cite{Bersin2019}. The strain field gradient, mainly distributed along the cantilever long axis due to the deformation (Fig.\ref{scheme}.(b)), shifts the optical transition frequency of each electron spin. By a well-designed strain field applied on the cantilever, we can shift the optical transition frequencies of NV centers to be different from each other and encode their location information with the transition frequencies (Fig.\ref{scheme}.(d)). This method can encode a large number of NV centers since the optical frequency shift ($\sim$ 100 GHz) can be much larger than its optical linewidth (<100 MHz). 

Besides individual readout, quantum operations require the precise control of every single spin. Our approach utilizes an external magnetic field gradient to split the optically detected magnetic resonance (ODMR) frequencies. The magnetic field and its gradient will be distributed along the cantilever long axis (Fig.\ref{scheme}.(a)). By carefully choosing the magnetic field gradient with respect to the position, the direction of the NV axis and the zero-field splitting of each spin due to the Zeeman effect is unique and enables the individual control without perturbing the others.

In order to realize multi-qubit quantum gates, we need to consider the coherence time of the spin. At 3.7 K, an electron spin coherence time longer than 1 second has been achieved for a NV center coupled to more than ten nuclear spins in a CVD-grown diamond with a natural 1.1\% abundance of $^{13}$C \cite{Abobeih2018}.   The longitudinal electron-spin relaxation time $T_1$ is about 3600~s at 3.7~K \cite{Abobeih2018}. 
Even for a HPHT (high pressure, high temperature)  diamond sample with a very high negatively charged NV concentration of 16 ppm (average NV-NV separation: 7 nm), the $T_1$  is measured to be about 50 ms at 100K \cite{Jarmola2012T1}.  They are long enough to realize schemes discussed in this paper. For example, an universal Toffoli gate for 3 qubits can be realized in just 50 $\mu$s.

In this paper, we describe the details of individual addressing and selective readout in the next section, which overcome the challenge of individual addressing without affecting nearby spins. 
Progress towards quantum computation in diamond requires a universal set of quantum gates and relatively long coherence time. We make use of the external driving to realize universal quantum gates in two-qubit and three-qubit system. The protocol and its operation error are demonstrated in the third section. However, this protocol limits the scalability of the system.
Alternatively,  we introduce the optimal control method on NV-NV system to design a more general approach to implement the quantum gates.

\section{Individual addressing}

When no external magnetic field is applied, $\left| m_s=1\right\rangle$ and $\left| m_s=-1\right\rangle$ states of the electron spin have identical energy if we neglect the hyperfine interaction. The ODMR frequencies of all NV centers are almost the same
making it hard to address a single spin. 
By virtue of the magnetic field gradient, we are able to split the ODMR frequencies and hence manipulate different NV centers independently by applying a global microwave (MW) pulse with different frequency components. 
The real ODMR linewidth will be broadened by the power of the MW field and other factors when conducting the electron spin resonance. Assuming a broadened ODMR linewidth of 100 kHz, an inhomogeneous magnetic field with a maximum value of $100$ G will enable us to individually manipulate more than $100$ NV centers by tuning the driving MW field. More NV centers can be individually manipulated if a larger magnetic field is used.

\begin{figure}[t]
 	\begin{center}
	\includegraphics[scale=0.5]{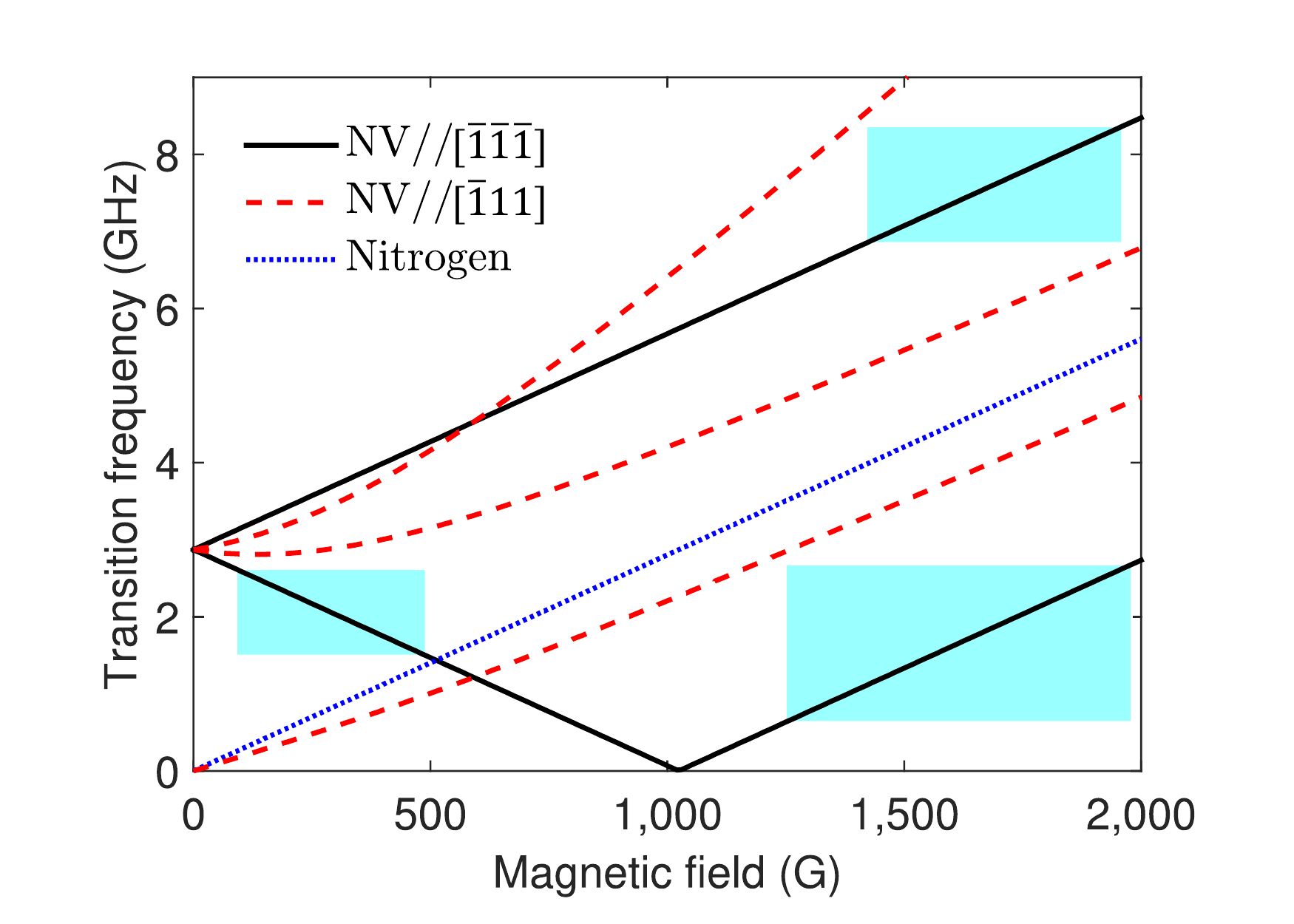}
	\caption{Transition frequencies between each sublevel of NV centers and nitrogen impurities as a function of the external magnetic field, which is along $\left[\bar1\bar1\bar1\right]$ axis. The black solid lines are transition frequencies between sublevels for NV centers aligned with the external magnetic field, i.e, the NV axis is along $\left[\bar1\bar1\bar1\right]$. They represent transitions from $\left| m_s=0\right\rangle$ to $\left| m_s=1\right\rangle$ and from $\left| m_s=0\right\rangle$ to $\left| m_s=-1\right\rangle$, from top to bottom, respectively. The red dashed lines are transition frequencies between sublevels for NV centers aligned at an angle of $109^\circ$ with respect to $\left[\bar1\bar1\bar1\right]$, i.e, NV-axis along $\left[\bar{1}11\right]$, $\left[1\bar{1}1\right]$ or $\left[11\bar{1}\right]$. From top to bottom, the transitions are from $\left| m_s=0\right\rangle$ to $\left| m_s=-1\right\rangle$, $\left| m_s=0\right\rangle$ to $\left| m_s=1\right\rangle$  and $\left| m_s=1\right\rangle$ to $\left| m_s=-1\right\rangle$, respectively. The blue dotted line stands for the transition from $\left| m_s=-1/2\right\rangle$  to $\left| m_s=1/2\right\rangle$  for nitrogen impurities. Three cyan shaded area stand for the applicable external magnetic field without cross relaxation and can be used to manipulate each NV center spin states individually.}
	\label{magnetic-field}
    \end{center}
\end{figure}

But beyond that, we also need to consider the uniqueness of the ODMR frequencies for each NV to avoid the crosstalk.
For a single NV center in the magnetic field of $\textbf{B}$, its Hamiltonian is $H_{NV}=DS_{z'}^2+g\mu_B \textbf{B}\cdot\textbf{S}$, where the direction of the magnetic field is along $z$ and $z'$ is the NV axis as shown in Fig.\ref{scheme}.(c).  The red sphere in Fig.\ref{scheme}.(c) denotes the nitrogen atom and the blue are for the carbon atoms. In the NV center's coordinates, the NV axis $z^\prime$ lies along the nitrogen-vacancy bond and can be one of four crystallographic directions: $[\bar{1}\bar{1}\bar{1}]$,$[\bar{1}11]$,$[1\bar{1}1]$ or $[11\bar{1}]$. $\textbf{S}$ and $S_{z'}$ are the spin vector and its projection along $z'$ direction of the electron spin.
$D=2.88$ GHz is the zero-field splitting frequency of the NV center. In the basis of the eigenvectors $\{\left|1\right\rangle, \left|0\right\rangle, \left|-1\right\rangle\}$ of $S_{z'}$ under the NV coordinates, the Hamiltonian can be expanded as\cite{Ma2017},
\begin{gather}
 H_{NV}=\begin{bmatrix}     
 D+\delta cos\theta &-\delta sin\theta\frac{e^{-i\phi}}{\sqrt{2}}&0\\
-\delta sin\theta\frac{e^{i\phi}}{\sqrt{2}}&0& -\delta sin\theta\frac{e^{-i\phi}}{\sqrt{2}}\\
 0& -\delta sin\theta\frac{e^{i\phi}}{\sqrt{2}}&D-\delta cos\theta
\end{bmatrix}
\end{gather}
where $\delta=g_e\mu_B B$ is the Zeeman shift when the NV axis aligns with the magnetic field, $g_e$ is the electronic g-factor, $\mu_B$ is the Bohr mangeton, $\theta$ is the angle between the NV-axis $z'$ and the magnetic field $z$. $\phi$ is the azimuthal angle of the NV axis in the $x-o-y$ plane in the cantilever coordinates. The eigenvalues of $H_{NV}$ and the transition frequencies are independent of $\phi$.
 The shifted transition frequencies between different energy levels as a function of the magnetic field are shown in Fig.\ref{magnetic-field}. Here the external magnetic field is aligned parallel to the $\left[\bar1\bar1\bar1\right]$ direction of the crystal diamond so $\theta=0^{\circ}$ is for those the NV axis aligned along $\left[\bar1\bar1\bar1\right]$ and $\theta=109^{\circ}$ is for those aligned along $\left[\bar{1}11\right]$, $\left[1\bar{1}1\right]$ and $\left[11\bar{1}\right]$.  
The black solid lines correspond to the transitions from $\left| m_s=0\right\rangle$ to $\left| m_s=1\right\rangle$ and from $\left| m_s=0\right\rangle$ to $\left| m_s=-1\right\rangle$ when $\theta=0^{\circ}$, from top to bottom, respectively.
For this kind of NV centers, the Zeeman shift is proportional to the external magnetic field magnitude.
For the case $\theta=109^{\circ}$, the magnetic field mixes the spin states so the non-spin-conserving transitions are also allowed. Three red dashed lines from top to bottom represent the transitions from $\left| m_s=0\right\rangle$ to $\left| m_s=-1\right\rangle$, from $\left| m_s=0\right\rangle$ to $\left| m_s=1\right\rangle$ and from $\left| m_s=1\right\rangle$ to $\left| m_s=-1\right\rangle$, respectively. 
Similar to the Zeeman effect on NV centers, the transition frequency from $\left| m_s=-1/2\right\rangle$ to $\left| m_s=1/2\right\rangle$ of nitrogen impurities as a function of the magnetic field is shown in the blue dotted line. 

\begin{figure*}
\begin{center}
	\includegraphics[scale=0.3]{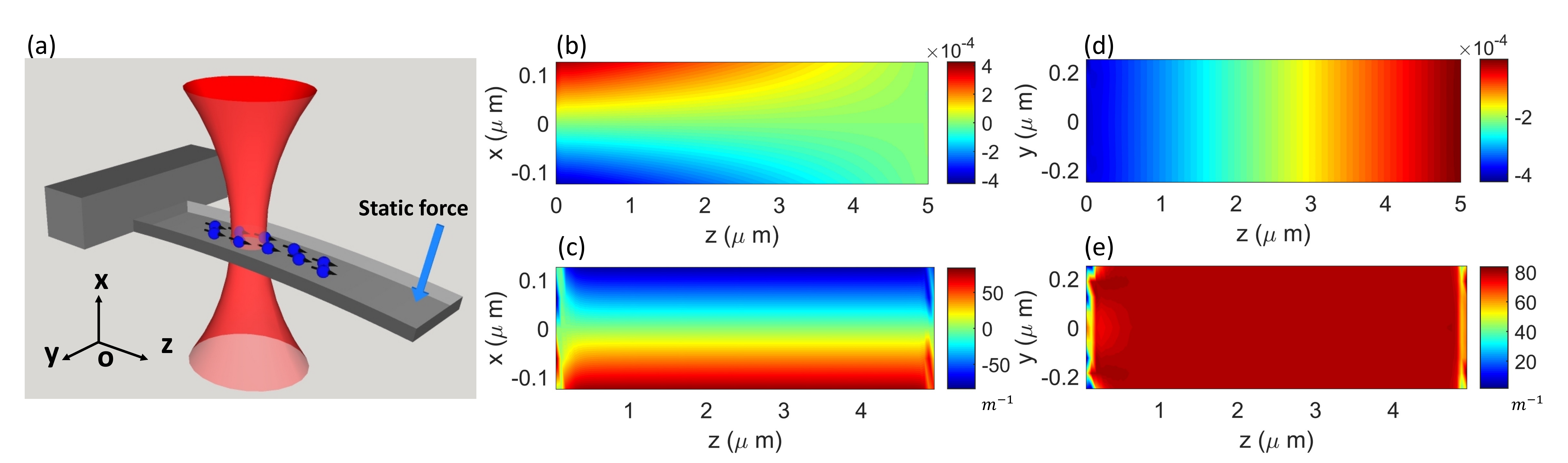}
	\caption{(a) Schematic of the diamond cantilever. The strain is applied along z axis in the cantilever coordinates, which is [$\bar1\bar1\bar1$] in Miller indices for diamonds. A red tunable laser is applied to a cluster of NV centers in the cantilever for individual control and readout. (b)-(e) The simulated distribution of strain and strain gradient of the diamond cantilever when it is bent by an external force. The cantilever is taken to be $5$ $\mu$m in length, $0.5$ $\mu$m in width and $0.25$ $\mu$m in height. The maximum value of the stress in the simulation is $487$ MPa, which is much smaller than its ultimate tensile strength. In such condition, the local strain on the cantilever can reach up to $10^{-4}$ .The side view  ($z-o-x$ plane) of the simulated distribution of strain is shown in (b) and of strain gradient is shown in (c). The top view  ($z-o-y$ plane) of the distribution of strain is shown in (d) and of strain gradient is shown in (e). }
	\label{cantilever-structure}
\end{center}
\end{figure*}

In order to manipulate the spin state of each NV center independently, without affecting the others, the range of the applied external field is required to be well designed. The most concerned part is the cross relaxation that happens when the transition frequencies between different spin states of different NV centers and the nitrogen impurities coincide\cite{Jarmola2012T1}. 
Such cross relaxation would strongly lower the spin-lattice relaxation time $T_1$\cite{Reynhardt:1998aa, Jarmola2012T1} and hence degrade the performance of the qubit. To protect the spin states and address a single NV center independently, the transition frequency between specific sublevels is required to be  monotonous as a function of the external magnetic field and has an unique value in the range without coinciding with other transitions. 
In Fig.\ref{magnetic-field}, three cyan shaded areas represent the applicable magnetic field excluding cross relaxation. We notice that the most achievable magnetic field tunable range is from $100$ G to $400$ G, i.e. the area in the bottom left-hand corner, which can shift the ODMR frequency by $0.8$ GHz. If the ODMR linewidth settles at $200$ kHz, corresponding to a coherence time $T_2^*\approx 2$ $\mu$s\cite{Rondin:2014aa}, the system will be able to individually manipulate more than 100 different NV electron spins, which is a promising technique for building large-scale quantum network.

Now we consider using a novel microscopy technique to optically read out individual NV centers in a closely-spaced ensemble. We apply a strain field gradient to shift the optical transition frequency of each NV center. Similar to the magnetic field,  the shift of the optical transition frequency due to the strain field is required to be different for each NV center.
If we apply a 637-nm linearly polarized  tunable laser on the sample constantly and take a confocal map simultaneously to monitor the NV photoluminescence, we will be able to get the ZPL of each individual NV centers as well as their positions. 
The excitation resonance method also enables us to read out the spin state of each NV without affecting others\cite{Bersin2019}. 

The strain field gradient will be realized by the stress of the cantilever delivered by the contact with an AFM tip or by an external static force . Here we discuss the details of the latter case.
The schematic of the diamond cantilever system is shown in Fig.\ref{cantilever-structure}.(a). The cantilever is fixed by one end and applied by an external static force on the other end. The bending leads to the inhomogeneous strain field on the cantilever. 

We first simulate the distribution of strain field and its gradient in the diamond cantilever system. In the simulation, the cantilever is set to be $5$ $\mu$m in length, $0.5$ $\mu$m in width and $0.25$ $\mu$m in height. 
An external force of $5\times 10^{-4}$ N is applied on one side. The maximum value of the stress in this situation is calculated to be $487$ MPa, which is one order smaller than the ultimate tensile strength $2.9$ GPa of bulk diamond. Fig.\ref{cantilever-structure}.(c)-(e) show the simulated distribution of strain field and its gradient from the side view (b),(c) and from the top view (d), (e). It is worth to notice that in this situation, the strain can reach up to $4\times 10^{-4}$, which is large enough to split the ZPL frequency for individual addressing. Besides, we mainly consider the NV centers that are near the surface of the diamond cantilever. Fig.\ref{cantilever-structure}.(c) and (e) shows that the gradient strain field is mainly along $z$ axis of the cantilever.
 This part will be discussed later in this section.

Next, we analyze the optical transition frequency as a function of the strain field magnitude for the NV centers with different orientations.  
The negatively charged NV center is known to consist of a ground triplet state $^{3}A_{2}$, an optical excited doublet-triplet state $^{3}E$ and several dark states\cite{Doherty:2012aa}.
The ground state$^{3}A_{2}$ and the excited state $^{3}E$  are associated with a ZPL energy around 637 nm (1.945 eV). 
If we take the spin-orbit interaction as well as the spin-spin interaction into consideration, the ground state $^{3}A_{2}$ is split into a spin single $m_s=0$ and a spin doublet $m_s=\pm1$ with the zero field splitting energy of $2.88$ GHz which can be identified by ESR, while the excited state  $^{3}E$ is split by several GHz \cite{Tamarat2008}.
The fine structures of the excited states are highly dependent of the strain and deformation, as well as the temperature. 
Some recent studies show that the measured frequency shift for the excited state is $1.39$ THz/GPa under hydrostatic pressures\cite{Lee2016strain}, while the ground states are weakly coupled to the strain. The pressure dependence for the ground state is measured to be $14.58$ MHz/GPa\cite{doherty2014pressure}. 
Hence, with the externally applied strain on the NV center, we can neglect the effect of strain on the ground state and only consider the energy shift of the excited state. 

 \begin{figure*}[t]
 \begin{center}
	\includegraphics[scale=0.35]{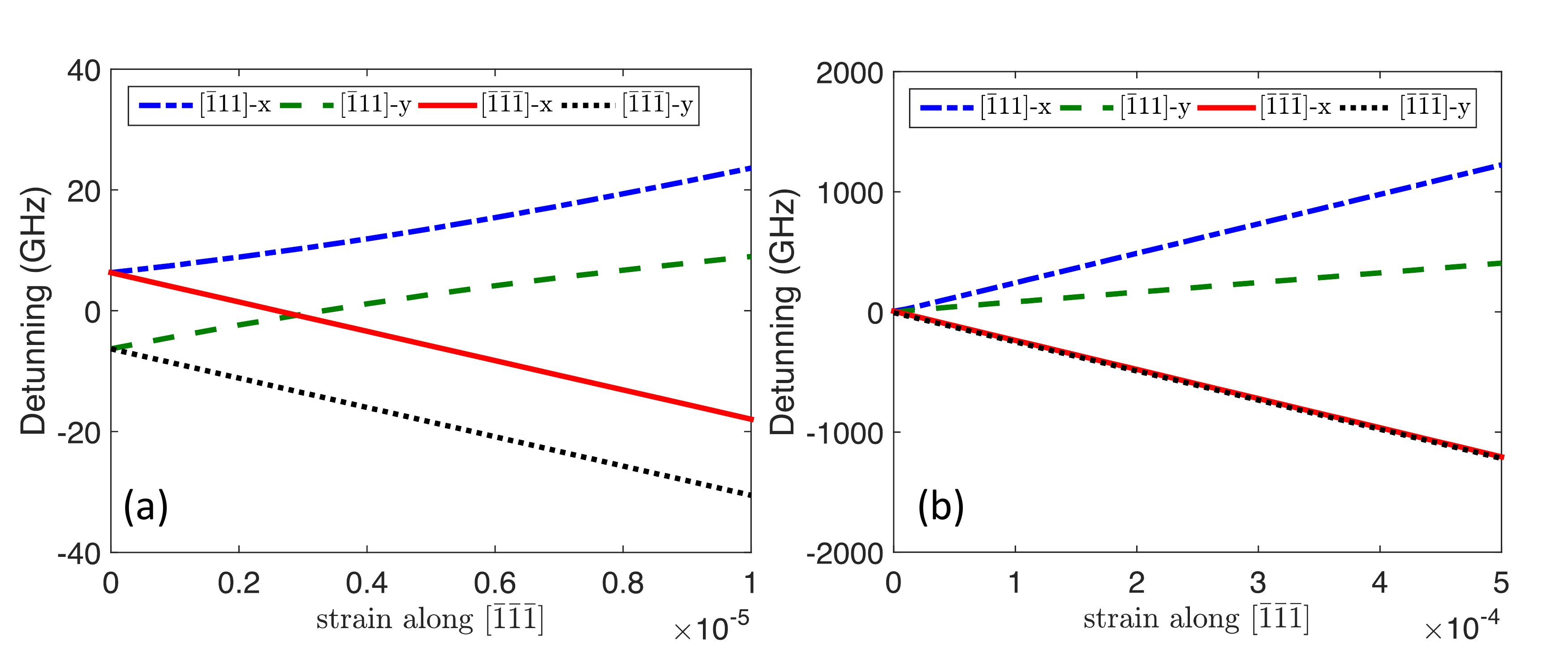}
	\caption{Detuning optical transition frequencies for NV centers with different orientations as a function of the applied strain field along $[\bar1\bar1\bar1]$. For NV centers with $\hat{z} $ axis of $[\bar111]$,  $[1\bar11]$ and $[11\bar1]$ have identical energy for $E_x$ and $E_y$, shown as the blue dashed-dotted line and the green dashed line, respectively. While for NV centers orientated along $[\bar1\bar1\bar1]$, the detuning frequencies for $E_x$ and $E_y$ are shown in the red solid line and the black dotted line, respectively. (a). For strain field in a range from $0$ to $1\times 10^{-5}$. (b) For a larger range from $0$ to $5\times 10^{-4}$.
	}
	\label{strain}
	\end{center}
\end{figure*}

Since we focus on the dependence of extra strain on the bright states of NV centers,  the orbital doublet $E_x$ and $E_y$ for $m_s=0$ are the only two sublevels of the excited state considered for the following discussion. The perpendicular component and the axial component of the strain field with respect to the NV axis have different affects on the energy shift. The perpendicular component shifts the optical transition frequency of $E_x$ and $E_y$ in an opposite way and lifts their degeneracy, while the axial component shifts the energy  uniformly\cite{Manson:2006aa}.

In order to split the optical transition frequencies of each NV center and address them individually, we need to study the dependence of frequency on the strain field first. Under the external strain, the frequencies of transitions from the ground state to the excited states, which are $A\rightarrow E_x$ and $A\rightarrow E_y$, can be expressed as \cite{Lee2016strain}.
\begin{eqnarray}
E_x=f_{ZPL}+g_{A_1}+\sqrt{(g_{E_1}+\delta f_{E1})^2+(g_{E_2}+\delta f_{E_2})^2},\nonumber\\
E_y=f_{ZPL}+g_{A_1}-\sqrt{(g_{E_1}+\delta f_{E1})^2+(g_{E_2}+\delta f_{E_2})^2},
\label{transition-frequency}
\end{eqnarray}
where $f_{ZPL}$ is the natural ZPL frequency, $\delta f_{E_1}$ and $\delta f_{E_2}$ represent the intrinsic strain field induced frequency shift of symmetry $E_1$ and $E_2$, respectively. From the previous work\cite{Tamarat2008,Tamarat2006linewidth}, $\delta f_{E_1}$ and $\delta f_{E_2}$ are around $6.3$ GHz and $0.15$ GHz, respectively. 
While $g_{A_1}$, $g_{E_1}$ and $g_{E_2}$ correspond to the frequency shift as a function of the external applied strain with symmetry $A_1$, $E_1$ and $E_2$ and they can be written as 
\begin{eqnarray}
g_{A_1}=\lambda_{A_1}\epsilon_{zz}+\lambda_{A_1\prime}(\epsilon_{xx}+\epsilon_{yy}),\nonumber\\
g_{E_1}=\lambda_{E}(\epsilon_{yy}-\epsilon_{xx})+\lambda_{E\prime}(\epsilon_{xz}+\epsilon_{zx}),\nonumber\\
g_{E_2}=\lambda_{E}(\epsilon_{xy}+\epsilon_{yx})+\lambda_{E\prime}(\epsilon_{yz}+\epsilon_{zy}),
\label{coupling-strength}
\end{eqnarray}
Here we use $\epsilon$ as the strain tensor under the NV center coordinates, while $\lambda_{A_1}=-1.95$ PHz, $\lambda_{A_1\prime}=2.16$ PHz, $\lambda_{E}=-0.85$ PHz and $\lambda_{E\prime}=0.02$ PHz are the parameters of orbital strain coupling \cite{Lee2016strain}.

As mentioned before, an external static force is applied at one end of the cantilever and hence a gradient field is distributed along z axis in the cantilever coordinates, which is [$\bar{1}\bar{1}\bar{1}$] in Miller index for diamond lattice. The cantilever geometry with the desired lattice structure can be realized by using focused ion beam (FIB) fabrication and tilting the bulk diamond to different angles\cite{Babinec:2011aa}. In this system, we first choose to use the strain tensor in the cantilever's coordinates $(x,y,z)$ as
 $\epsilon_c=[  \{-\nu\epsilon , 0 , 0\}, \{0,-\nu\epsilon,0\}, \{0,0 ,\epsilon\}]$, 
where $\epsilon$ is the mechanically induced strain along z direction and $\nu=0.11$ is the Poisson ratio of diamond. 
  
In order to analyze frequency shifts due to the strain field for four possible NV axes, we need to introduce NV centers' coordinates $(x', y', z')$. It is known that $z^\prime$ axis is identified as the NV axis. While the NV $x^\prime$ axis is chosen to lie along the projection of one of the the carbon bond in the plane perpendicular to the NV axis.
We transform the above strain tensor from the cantilever coordinates $(x, y, z)$ to the NV center coordinates $(x^\prime, y^\prime, z^\prime)$ and use Eq.\ref{transition-frequency} to get the optical transition frequencies between each sublevels.

 For the NV centers with $z'$ axis along $[\bar{1}11]$ direction, the transformed strain tensor is 
 \begin{gather}
\epsilon_{\bar{1}11}=
  \begin{bmatrix}
    \frac{2}{9}-\frac{7}{9}\nu & \frac{2\sqrt{3}}{9}(\nu+1) & \frac{\sqrt{2}}{9}(\nu+1) \\
    \frac{2\sqrt{3}}{9}(\nu+1) & \frac{2}{3}-\frac{1}{3}\nu & \frac{\sqrt{6}}{9}(\nu+1)  \\
     \frac{\sqrt{2}}{9}(\nu+1) & \frac{\sqrt{6}}{9}(\nu+1) & \frac{1}{9}-\frac{8}{9}\nu\\
  \end{bmatrix}\epsilon.
  \end{gather}
The NV centers with $[1\bar{1}1]$ and $[11\bar{1}]$ orientation have the identical energies of $E_x$ and $E_y$ as above. 
When the NV axis is aligned along $[\bar{1}\bar{1}\bar{1}]$ direction, the transformed strain tensor becomes
\begin{gather}
\epsilon_{\bar{1}\bar{1}\bar{1}}=\epsilon_c
\end{gather}

The calculated detuning optical transition frequencies for NV centers with different orientations as a function of the strain field are shown in Fig.\ref{strain}. Here we only focus on the detuning part so the absolute optical transition frequencies should be $f=f_{Detuning}+f_{ZPL}$ , where $f_{ZPL}=470.6$ THz is the ZPL frequency of NV center. 
For NV centers orientated along $[\bar{1}11]$, $[1\bar{1}1]$ and $[11\bar{1}]$, the detuning energy for $E_x$ and $E_y$ are shown in the blue dashed-dotted line and the green dashed line in Fig.\ref{strain}.(a), respectively. While the red solid line and the black dotted line show the detuning energy of $E_x$ and $E_y$ for those aligned along $[\bar1\bar1\bar1]$. 
The detuning frequencies for different orientations and sublevels as a function of the strain field in a larger range are shown in Fig.\ref{strain}.(b). 
We notice that for NV centers orientated along $[\bar1\bar1\bar1]$, the optical transition frequencies can be shifted by $24$ GHz under the strain of $10^{-5}$. 
The linewidth of optical transition frequency is around $\Gamma_0/2\pi=13$ MHz at low temperature. Thus more than $100$ NV centers can be addressed individually by this approach.

\section{Universal quantum gates }

 \begin{figure*}[t]
 \begin{center}
	\includegraphics[scale=0.5]{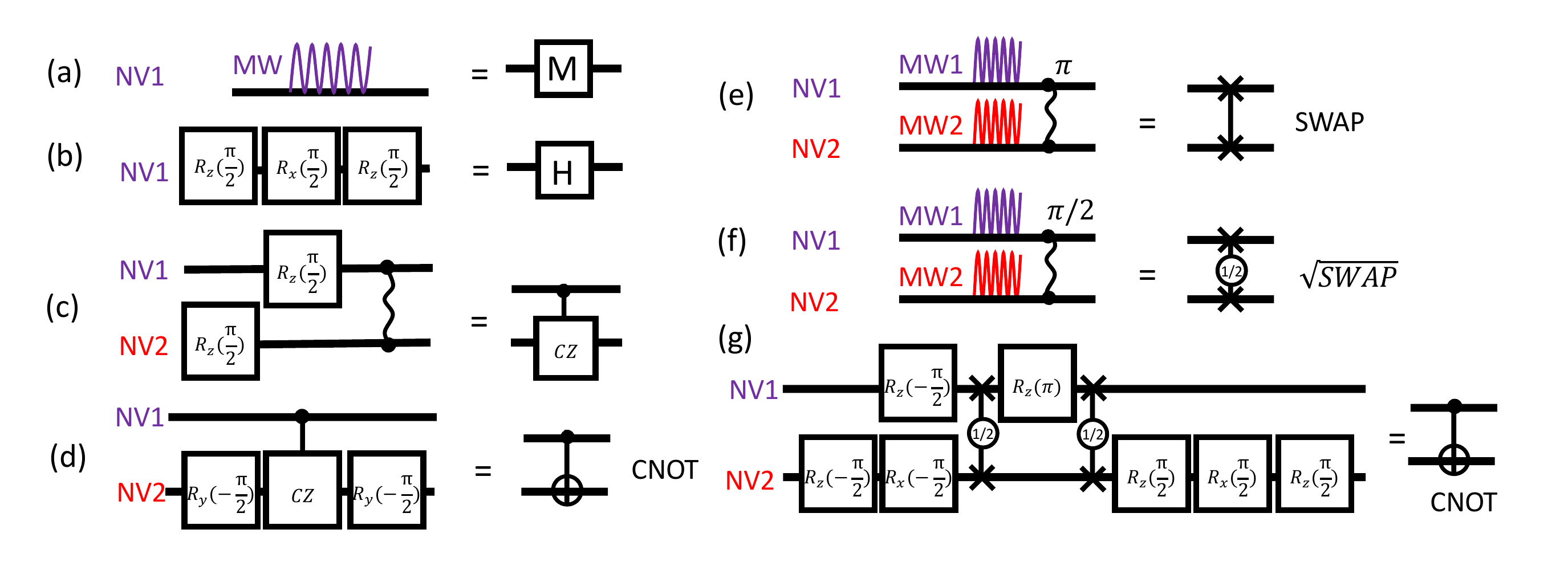}
	\caption{Schematic for the quantum gates of single-qubit and double-qubit system in diamond. (a) Arbitrary rotation gate $M$ on a single qubit by applying resonant MW pulse and tuning the duration and phase of the pulse. (b) Hadamard gate on a single qubit by three rotations of $\pi/2$. (c).The controlled-Z (CZ) gate on two qubits by two rotations on single qubit and their ZZ interaction. (d) The controlled-NOT gate by a CZ gate and two rotations on single qubit. (e) Under two simultaneous resonant MW pulses on two qubits, XX interaction between two qubits is turned on. The SWAP gate is realized when the duration of the XX interaction is $2/\nu_{dip}$. (f) The square root of SWAP (denoted by $\sqrt{SWAP}$) can be obtained by changing the duration time to be able $1/\nu_{dip}$. (g) The CNOT gate is obtained by two $\sqrt{SWAP}$ gates and seven single-qubit rotations. 
	}
	\label{gate}
	\end{center}
\end{figure*}

 \begin{figure*}[tbp]
 \begin{center}
	\includegraphics[scale=0.3]{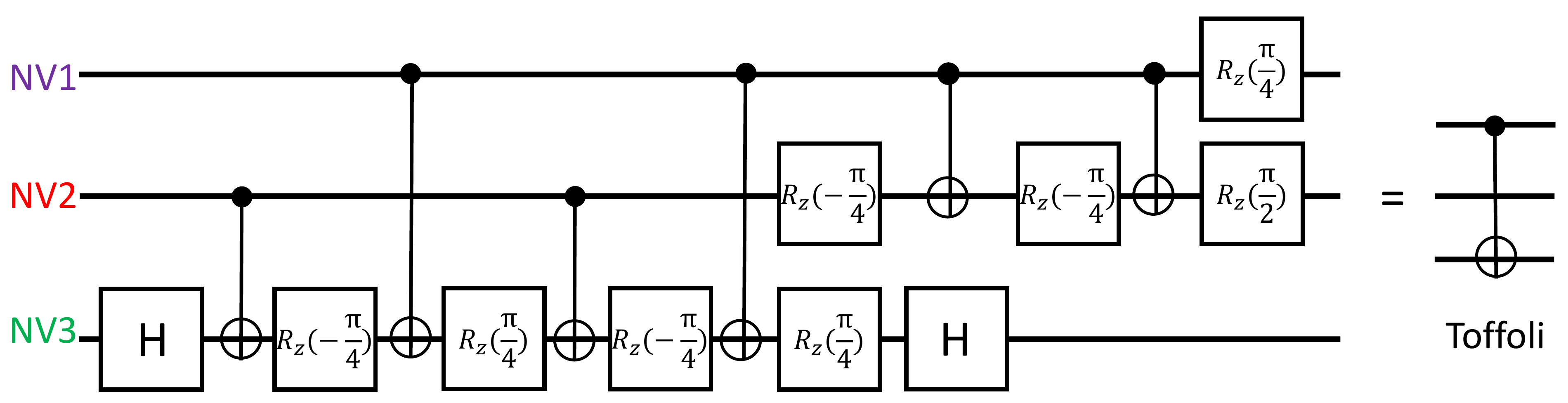}
	\caption{Schematic for the Toffoli gate of three-qubit system in diamond. 
	}
	\label{gate_3}
\end{center}
\end{figure*}

Now we discuss how to create universal quantum gates for multiple closely-spaced NV centers  for quantum information processing.  
The entanglement among three qubits will be realized by the spin-spin dipolar coupling between each NV centers. 
The system of three NV centers with mutual interaction is described by the Hamiltonian,
\begin{eqnarray}
H=H_{A}+H_{B}+H_{C}+H^{dip}_{AB}+H^{dip}_{AC}+H^{dip}_{BC},
\end{eqnarray}
where $H_{i}=D\cdot S_{zi}^2+g_{e}\mu_B\vec{B}\cdot\vec{S_{i}}$ stands for the self Hamiltonian of the $i^{th}$ single NV center, $D=2.88$ GHz is the zero-field splitting  and $\gamma_e=g_e\mu_B=-2.8$ MHz/Gauss is the electronic spin gyromagnetic ratio. 
The NV-NV electron dipolar interaction is given by 
\begin{eqnarray}
H^{dip}_{AB}&=\frac{\mu_{0}}{4\pi}\frac{g_e^2\mu_B^2}{r_{AB}^3}[S_{A}\cdot S_{B}-3(\vec{S_{A}}\cdot \vec{n_{AB}})(\vec{S_{B}\cdot\vec{n_{AB}}})]\notag\\
& \approx \nu_{dip}S_{zA}S_{zB},
\end{eqnarray}
where $\mu_0$ is the magnetic permeability,  $r_{AB}$ is the distance between spin A and B, $\vec{n_{AB}}$ is the unit vector from A to B and here we set $h=1$.
We have discarded the spin flip-flop terms due to its small strength around $100$ kHz, negligible compared to the spin-dependent energy difference of $2.88$ GHz. When two non-parallel NV electron spins are separated by $8$ nm, the measured dipolar coupling strength is $\nu_{dip}=42.7$ kHz\cite{Neumann:2010ab}, while $\nu_{dip}=101$ kHz for parallel electron spin pairs with the same separation. 
The spin coherence time $T_2$ can be more than $1$ s in a CVD-grown single-crystal diamond at 3.7K \cite{Abobeih2018}.
Therefore, the dipolar coupling between two NV centers can be used to coherently couple multiple qubits.

We first consider the two-qubit Hamiltonian $H=H_{sys}+H_{mw}$ where $H_{sys}=\omega_1S_{z1}+\omega_2S_{z2}+\nu_{vip}S_{z1}S_{z2}$ has the form of the Ising model. 
A universal set of two-qubit operations should consist of arbitrary single-qubit rotation and a controlled-NOT (CNOT) gate. 
Here we consider the $m_s=0$ and $m_s=-1$ states of the electron spins as $\left|0\right\rangle$ and $\left|1\right\rangle$ of the qubit, respectively. 
When a large MW field with a certain frequency is applied to the qubits, we can make the approximation that $e^{-iHt/\hbar}\approx e^{-iH_{mw}t}$\cite{Nielsen:2010aa}. 
In this way, we can achieve arbitrary single-qubit operations $R_{c}(\theta)=e^{-iC\theta/2}$, i.e. rotation around $c=x,y,z$ axis for any arbitrary angle as shown in Fig.\ref{gate}.(a), by changing the duration and the phase of the resonant MW pulse. The carrier frequency is taken to be the transition frequency $\omega_i$ between $m_s=0$ and $m_s=-1$, including the zero-field splitting, the Zeeman shift and the hyperfine interaction.

Besides the single-qubit rotation, the CNOT gate is an essential quantum gate for constructing quantum computers. 
It is represented in the process density form as 
$U_{CNOT}=  [\{1,0,0,0\} ,\{0,1,0,0\} ,\{0,0,0,1\},\{0,0,1,0\} ]$
in the computational basis $\{\left|00\right\rangle, \left|01\right\rangle, \left|10\right\rangle, \left|11\right\rangle\}$.
Hence, in this two interacting NV system, the CNOT gate is aimed at flipping one electron spin if the other electron spin is in the $\left|1\right\rangle$ state, otherwise it will remain in the same state.

The dipolar coupling strength between two NVs is tens  of kilohertz at 10 nm separation, which is far smaller compared to the energy difference (many megahertz) between quantum states. The CNOT gate can not be realized by the direct coupling for its small strength so external driving has to be implemented to achieve the quantum gates.

Since the coherent coupling between two NV centers is too weak to contribute to a direct spin flip-flop, the direct way to realize CNOT gate in two-qubit system is through the phase accumulation of ZZ interaction and the single-qubit gate as follows\cite{Dolde:2013aa}\cite{Dolde:2014aa},
\begin{equation}
U_{CNOT}=H_{1}U_{CZ}H_{1}
\end{equation}
where $H_{1}$ is the Hadamard gate for NV1 while remaining NV2 unchanged and $U_{CZ}$ is the controlled-Z (CZ) gate written as $U_{CZ}=  [\{1,0,0,0\},\{ 0,1,0,0 \},\{ 0,0,1,0\},\{0,0,0,-1\}]$, which can be realized by 
\begin{equation}
U_{CZ}=e^{i\pi/4}e^{iS_{z1}S_{z2}\pi/4}e^{-iS_{z1}\pi/4}e^{-iS_{z2}\pi/4}.
\end{equation}
The schematic of the CZ gate and the CNOT gate are shown in Fig.\ref{gate}.(c) and (d), respectively.

A scalable system needs the implement of parametrical modulation of qubit frequencies to realize entanglement and quantum gates among multiple electron spins\cite{2018arXiv180603886L,Yao:2011aa}. In this method, MW driving $H_{drive}=\Sigma_{i=1,2}\Omega_i cos(\omega_i t+\phi_i)S_i^z$ with different resonant frequencies are applied to the system to effectively modify the energy of each spin. Under the rotating wave approximation, the system Hamilton is rewritten as Jaynes-Cummings model in the rotated basis\cite{Yao:2011aa}, 
\begin{equation}
H_{int}=\frac{\nu_{vip}}{4}(S_{1}^{+}S_{2}^{-}+S_{1}^{-}S_{2}^{+})+\Omega_1S_{1}^z+\Omega_2S_{2}^z,
\end{equation}
where $\Omega_i$, $\omega_i$ and $\phi_i$ are the Rabi frequency, the carrier frequency and the phase of the external driving applied on the $i$th spin, respectively. By simply controlling the duration of the interaction to be $2/\nu_{dip}$, the quantum gates are swapped, as shown in Fig.\ref{gate}.(e). To realize the universal operation CNOT gate, we need to utilize the square root of the SWAP gate (denoted as a $\sqrt{SWAP}$) instead of the SWAP gate itself.   The process is similar to the SWAP gate except that the duration time of both MW pulses becomes $1/\nu_{dip}$.
Combined with seven single-qubit $\pi/2$ and $\pi$ pulses, the CNOT gate is achieved shown in Fig.\ref{gate}.(g). Therefore, the external driving enables us to selectively tune the interaction by turning on the MW on both spins and this approach applies to multi-qubit system.

For a three-qubit system, the Toffoli gate is a quantum logic gate which can universally realize both classical reversible computation and quantum computation\cite{Yu:2013aa}. 
It can be realized by 
\begin{eqnarray}
U_{Toffoli}=H_3U_{CNOT}^{23}R_z^3(-\frac{\pi}{4})U_{CNOT}^{13}R_z^3(\frac{\pi}{4})U_{CNOT}^{23}\nonumber\\
R_z^3(-\frac{\pi}{4})R_Z^2(\frac{\pi}{4})H_3U_{CNOT}^{12}R_z^2(-\frac{\pi}{4})U_{CNOT}^{12}R_z^2(\frac{\pi}{2})R_z^1(\frac{\pi}{4}).
\end{eqnarray}
The schematic of realizing a Toffoli gate by ten single-qubit gates and six CNOT gates is shown in Fig.\ref{gate_3}.

The other thing we need to consider is the performance of such quantum operations. It has always been a challenge to fulfill entanglement between spins while maintain high fidelity during the operation process due to inevitable noise in the system. As the analysis in \cite{NYao2012scalable}, the error probability or the infidelity of quantum operations from different noise sources is written as\
\begin{eqnarray}
P_{err}=P_{T_1}+P_{T_2}+P_{mw}+P_{mag}+P_{str}+P_{dip}\nonumber\\
= \frac{t}{T_1}+\frac{t^3}{T_2^3}+\frac{\delta_1^2}{\Omega_{mw}^2}+\frac{\Omega_{mw}^2}{\Delta_{mag}^2}+\frac{\Omega_{opt}^2}{\Delta_{str}^2}+\frac{\nu_{dip}^2}{\Omega_{mw}^2},
\label{eq:error}
\end{eqnarray}

The first and the second term correspond to the depolarization error and the dephasing error of a single NV center, respectively. They are determined by the operation time $t$ as well as the spin-lattice relaxation time $T_1$ or the spin coherence time $T_2$.
The third term is the crosstalk from the imperfections in the MW pulse operations. The error is represented by the fluctuation of the control field strength $\delta_1$ and the Rabi frequency of the MW pulse $\Omega_{mw}$.
The fourth term comes from the off-resonant crosstalk induced by the external magnetic field gradient for spin state control between two NV centers. The error depends on the Rabi frequency of the MW pulse $\Omega_{mw}$ and the splitting energy due to the magnetic field gradient $\Delta_{{mag}}$.
The fifth term is similar to the fourth one and it depends on the Rabi frequency of the optical excitation $\Omega_{opt}$ and the detuning $\Delta_{str}$ due to the strain field gradient.
The last term  corresponds to off-resonant error induced by the dipolar interaction between two NV centers and it  can be  represented as $(\nu_{dip}/\Omega_{mw})^2$.

Taking the following conservative numbers for NV centers in a CVD-grown diamond at 4~K \cite{Abobeih2018,Jarmola2012T1}: $T_1=50$~ms,  $T_2=1$~ms,  $\delta_1= 10$ kHz, $\Omega_{mw}=800$ kHz, $\Omega_{opt}=10$ MHz, $\Delta_{mag}=20$ MHz, $\Delta_{str}=500$ MHz, $\nu_{dip}=100$ kHz, the errors are $P_{T_1}=4\times10^{-4}$, $P_{T_2}=8\times10^{-6}$, $P_{mw}=1.56\times 10^{-4}$, $P_{mag}=1.6\times 10^{-3}$, $P_{str}=4\times 10^{-4}$, and $P_{dip}=1.95\times 10^{-3}$ for each term if the total operation time $t=20$ $\mu$s. The total error probability $4.5\times 10^{-3}$ is quite small.

Noticeably, the second term $P_{T_2}={t^3}/{T_2^3}$ in error probability strongly depends on the total operation time $t$ and the coherence time $T_2$. 
To finish the spin flip of the SWAP gate and then obtain the CNOT gate, the entire process takes at least two complete evolution period of a spin flip, which is $2/\nu_{vip}\approx3$ $\mu$s. 
For three-qubit Toffoli gate, on the hand, the whole operation needs six CNOT gates, which takes at least twelve complete evolution period. In this way, the operation time $t$ is at least $18$ $\mu$s for three-qubit system, which is similar to the calculation previously.

In summary, we can achieve the universal quantum gates for two-qubit, three-qubit or system with more qubits by external driving on individual spins. However, this approach requires the spin coherence time to be much longer than the operation time. As the number of qubits increases, the operation time of a multi-qubit gate will increase rapidly, which is a big challenge. Moreover, the noise due to surrounding nuclear spin bath as well as the imperfection of control pulse will degrade the gate fidelity.  In the following section, we discuss the optimal control method that can create universal quantum gates in multi-qubit   systems with high fidelity. 

\begin{figure*}[t]
 \begin{center}
	\includegraphics[width=1\textwidth]{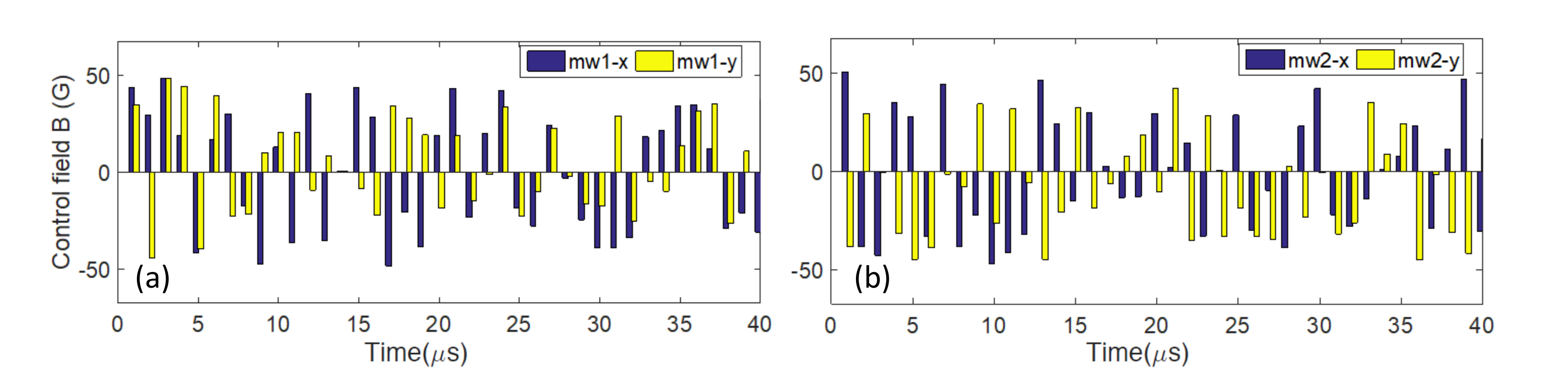}
	\caption{Optimal control for the CNOT gate for two interacting NV electron spins, with the spin-dependent splitting of $2700$ MHz and $2600$ MHz, respectively. The dipolar coupling strength between two NVs is taken to be $100$kHz. The sequence consists of 40 rectangular independent pulses with the total duration time of 40 $\mu$s.  
	Each pulse consists of four MW controls, with the frequency of the transition frequency between the $m_s=0\rightarrow m_s=-1$ transitions for two NV electron spins and the direction along $x$ and $y$ axis, respectively.  All four control fields are applied to the NV-NV system simultaneously during the whole pulse sequence. 
	}
	\label{OC_CNOT_2qubit}
\end{center}
\end{figure*}

\begin{figure*}
  \centering
      \includegraphics[width=0.9\textwidth]{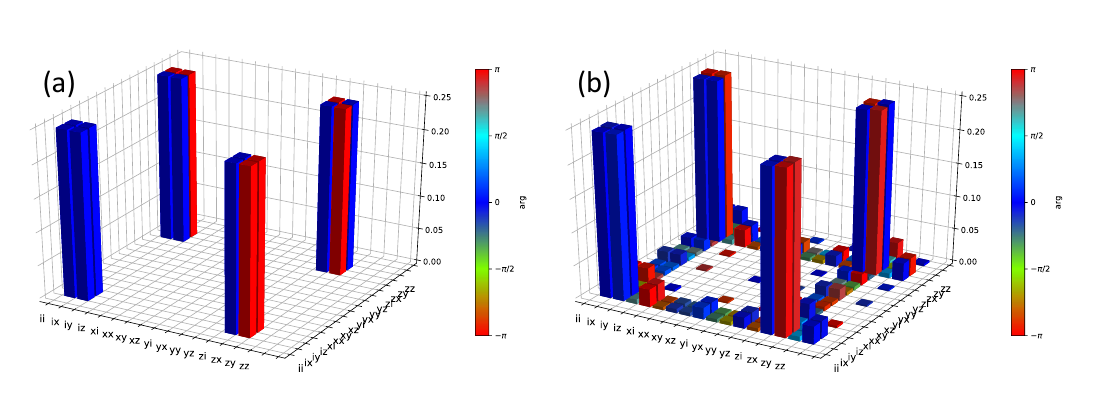}
  \caption{Process matrix of CNOT gate for two interacting NV center electron spins. (a) is the process matrix of the ideal CNOT gate in the computation basis defined by tensor products of Pauli operators $\{I, X, Y, Z\}$. (b) is the process matrix of the CNOT gate realized by optimal control shown above (F=0.9914). The bar height and the color show the absolute value and the phase of matrix elements in complex numbers form, respectively.}
  	\label{tomography}
\end{figure*}

\section{Optimal control on NV-NV system}
Optimal control theory is a general approach to manipulate the system dynamics by determining the control field and minimizing a cost functional. The control field consists of N piece of constant pulses, with different amplitude and phase of each piece\cite{Machnes:2011aa}. By optimizing the amplitude and the phase of the pulse sequence, high-fidelity entangled states\cite{Dolde:2014aa}, quantum gates\cite{Rong:2015aa} and quantum error correction\cite{Waldherr:2014aa} are achieved. In our own case, we utilize the gradient ascent pulse engineering (GRAPE) algorithm to design the quantum optimal control\cite{Khaneja:2005aa} in order to achieve high-fidelity universal quantum gates, even in the presence of noise by careful design. 
The main goal of optimal control is to maximize the gate fidelity, which is defined as the overlap between a quantum operator $U$ and a target unitary quantum gate $U_{ideal}$ as follows\cite{Nielsen:2002aa}\cite{Bowdrey:2002aa},
\begin{eqnarray}
F= \frac{1}{d(d+1)} (tr(MM^+)+\left | tr(M)\right |^2),
\end{eqnarray}
where $M=U^+_{ideal}U$ and $d/2$ is the dimension of the Hilbert space. Here $U$ is the overall quantum operation by all applied pulses in the sequence. 

 \begin{figure*}[t]
	\includegraphics[width=1\textwidth]{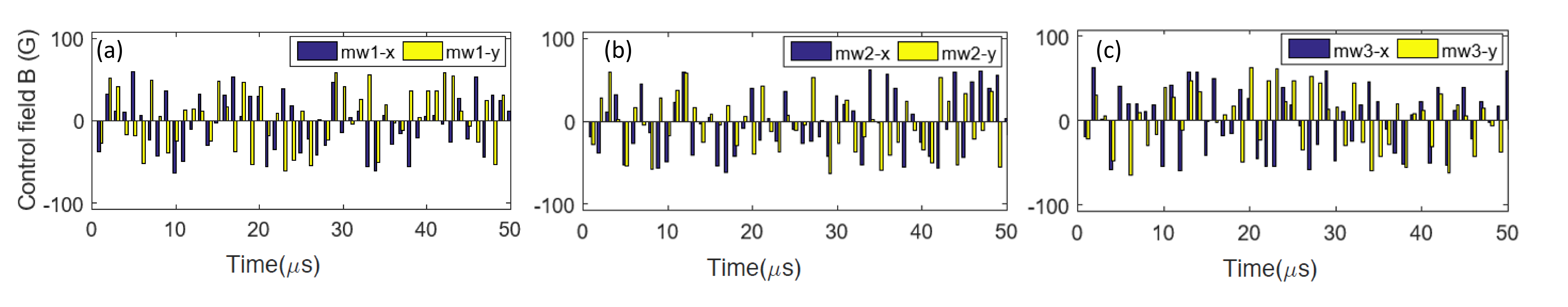}
	\caption{Optimal control for the Toffoli gate for three mutually interacting NV electron spins, with the spin-dependent splitting of $2700$ MHz, $2600$ MHz and $2500$ MHz, respectively. The dipolar coupling strengths between each two NVs are all identical be $100$kHz. The sequence consists of 50 rectangular independent pulse of 1 $\mu$s.  Each pulse consists of six control field components and they correspond to the transition frequencies of NV1, NV2 and NV3 along $x$ and $y$ axis. All six MW controls are applied to the three NVs system simultaneously during the whole pulse sequence. 
	}
	\label{OC_3qubit}
\end{figure*}

\begin{figure*}
  \centering
      \includegraphics[width=0.9\textwidth]{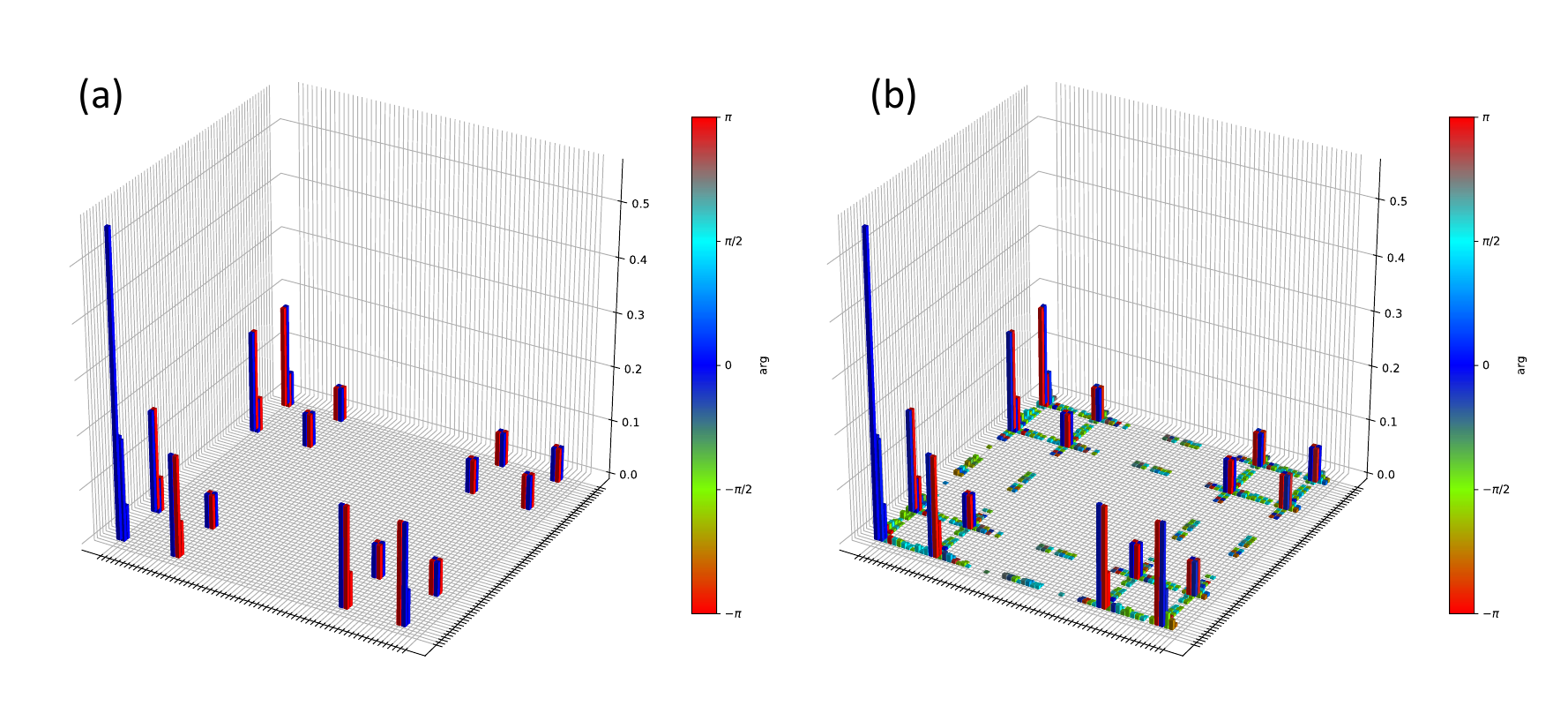}
  \caption{Process matrix of Toffoli gate for three mutually interacting NV centers. (a) is the process matrix of the ideal Toffoli gate in the computation basis defined by tensor products of Pauli operators $\{I, X, Y, Z\}$, of 64 elements. (b) is the process matrix of the Toffoli gate realized by optimal control shown above (F=0.9934). The bar height and the color show the absolute value and the phase of matrix elements in complex numbers form, respectively.}
  	\label{tomography_3qubit}
\end{figure*}

To characterize the effect of the control field, we use a  control Hamiltonian to represent as follows\cite{Rong:2015aa},
\begin{eqnarray}
H_c=-\gamma_e\Sigma_{i} cos[\omega_i t+\phi_i{t}]\textbf{B}_i(t)\cdot(\textbf{S}_1+\textbf{S}_2),
\end{eqnarray}
where $i=1, 2$ is the index for the NV center, $\omega_i$ is the carrier frequency of the control fields. Here $\omega_1$ and $\omega_2$ are taken to be $2700$ MHz and $2600$ MHz, as the transition frequencies between $\left|0\right\rangle$ and $\left|1\right\rangle$ of the two NV electron spins, respectively. $\textbf{B}_i$ is the magnetic field with frequency of $\omega_{i}$ and phase of $\phi_i$ applied on both qubits simultaneously. 
Alternatively, under the rotation frame approximation, the control field can be written as
\begin{eqnarray}
H_c=-\gamma_eB_0(u_{1x}(t)\cdot S_{1x}+u_{2x}(t)\cdot S_{2x}\nonumber\\
+u_{1y}(t)\cdot S_{1y}+u_{2y}(t)\cdot S_{2y}),
\end{eqnarray}
where we use $u_{1x}(t)$, $u_{1y}(t)$, $u_{2x}(t)$, $u_{2y}(t)$ to encode the amplitude and phase information of the MW pulse and they are changing independently and concurrently.
The designed optimal control sequence for the CNOT gate is shown in Fig.\ref{OC_CNOT_2qubit}. The sequence consists of 40 rectangular independent pulses with the total duration time of 40 $\mu$s. Each pulse consists of four MW controls, with the frequency identical with the transition frequency of NV1 (shown in Fig.\ref{OC_CNOT_2qubit}.(a)) and NV2 (shown in Fig.\ref{OC_CNOT_2qubit}.(b)) along $x$ and $y$ axis. 
All four control fields are applied to the NV-NV system simultaneously. Here the dipolar coupling strength between two spins is taken to be $100$ kHz. A gate fidelity of 0.9914 is achieved by using this control sequence.

The process matrices of the CNOT gate  realized by the optimal control sequence above are shown in Fig.\ref{tomography}.(b). They are represented  in the computation basis defined by the tensor products of Pauli operators $\{I,X,Y,Z\}$. The bar height and the color correspond to the absolute value and the phase of the matrix elements in complex numbers form, respectively. For comparison, the process matrices of the ideal CNTO gate is shown in Fig.\ref{tomography}.(a).

To realize a universal set of quantum operations on three-qubit system, we also study the optimal control method to realize the Toffoli gate. 
 The designed optimal control pulse sequence with a fidelity of 0.9934 and the tomography are shown in Fig.\ref{OC_3qubit}, \ref{tomography_3qubit}. The sequence consists of 50 rectangular independent pulses of $1$ $\mu$s for each.
Each pulse consists of six control field components and they correspond to the transition frequencies of NV1, NV2 and NV3 along $x$ and $y$ axis. In the design, the transition frequencies are set to be $2700$ MHz, $2600$ MHz and $2500$ MHz for three NVs and the coupling strength between all three are $100$ kHz.

Compared to the approach by using external driving introduced before, the optimal control method has the advantage of using shorter pulse duration to realize high-fidelity quantum operations. Noticeably, high fidelity of the quantum gates in NV-NV system can be realized in the presence of noise by adding the quasi-static noises from surrounding environment and imperfections of control field into the design\cite{Rong:2015aa}.

\section{Conclusion}

In this paper, we propose methods to use high-density diamond NV centers with an average separation on the order of $10$~nm for quantum computing. With the help of a strain gradient as well as a magnetic field gradient, we will be able to individually control and read out the single NV center. Combined with the optimal control method, high-fidelity universal quantum gates in a two-qubit and a three-qubit system are designed.
Scalability is the last consideration in the DiVincenzo criteria, which requires the system to combine the manipulations of individual qubits in a system that consist of a large number of qubits\cite{Loss:1998aa}. The narrow linewidths and the dispersive distributions of the optical transition frequencies and ODMR frequencies promise the individual control and read out of each NV in a cluster of more than 100 closely-spaced NV centers. The optimal control method can help to achieve high-fidelity quantum operations while minimize the operation time and protect the spin coherence of the system.

\section{Acknowledgements}
We thank  J. Ahn, and J. Bang for helpful discussions. This work is partly supported by the Purdue
Research Foundation, Tellabs Foundation and the NSF under Grant No. PHY-1555035.



\begin{thebibliography}{49}%
\makeatletter
\providecommand \@ifxundefined [1]{%
 \@ifx{#1\undefined}
}%
\providecommand \@ifnum [1]{%
 \ifnum #1\expandafter \@firstoftwo
 \else \expandafter \@secondoftwo
 \fi
}%
\providecommand \@ifx [1]{%
 \ifx #1\expandafter \@firstoftwo
 \else \expandafter \@secondoftwo
 \fi
}%
\providecommand \natexlab [1]{#1}%
\providecommand \enquote  [1]{``#1''}%
\providecommand \bibnamefont  [1]{#1}%
\providecommand \bibfnamefont [1]{#1}%
\providecommand \citenamefont [1]{#1}%
\providecommand \href@noop [0]{\@secondoftwo}%
\providecommand \href [0]{\begingroup \@sanitize@url \@href}%
\providecommand \@href[1]{\@@startlink{#1}\@@href}%
\providecommand \@@href[1]{\endgroup#1\@@endlink}%
\providecommand \@sanitize@url [0]{\catcode `\\12\catcode `\$12\catcode
  `\&12\catcode `\#12\catcode `\^12\catcode `\_12\catcode `\%12\relax}%
\providecommand \@@startlink[1]{}%
\providecommand \@@endlink[0]{}%
\providecommand \url  [0]{\begingroup\@sanitize@url \@url }%
\providecommand \@url [1]{\endgroup\@href {#1}{\urlprefix }}%
\providecommand \urlprefix  [0]{URL }%
\providecommand \Eprint [0]{\href }%
\providecommand \doibase [0]{http://dx.doi.org/}%
\providecommand \selectlanguage [0]{\@gobble}%
\providecommand \bibinfo  [0]{\@secondoftwo}%
\providecommand \bibfield  [0]{\@secondoftwo}%
\providecommand \translation [1]{[#1]}%
\providecommand \BibitemOpen [0]{}%
\providecommand \bibitemStop [0]{}%
\providecommand \bibitemNoStop [0]{.\EOS\space}%
\providecommand \EOS [0]{\spacefactor3000\relax}%
\providecommand \BibitemShut  [1]{\csname bibitem#1\endcsname}%
\let\auto@bib@innerbib\@empty
\bibitem [{\citenamefont {Matthews}\ \emph {et~al.}(2009)\citenamefont
  {Matthews}, \citenamefont {Politi}, \citenamefont {Stefanov},\ and\
  \citenamefont {O'Brien}}]{Matthews:2009aa}%
  \BibitemOpen
  \bibfield  {author} {\bibinfo {author} {\bibfnamefont {J.~C.~F.}\
  \bibnamefont {Matthews}}, \bibinfo {author} {\bibfnamefont {A.}~\bibnamefont
  {Politi}}, \bibinfo {author} {\bibfnamefont {A.}~\bibnamefont {Stefanov}}, \
  and\ \bibinfo {author} {\bibfnamefont {J.~L.}\ \bibnamefont {O'Brien}},\
  }\href {http://dx.doi.org/10.1038/nphoton.2009.93} {\bibfield  {journal}
  {\bibinfo  {journal} {Nature Photonics}\ }\textbf {\bibinfo {volume} {3}},\
  \bibinfo {pages} {346} (\bibinfo {year} {2009})}\BibitemShut {NoStop}%
\bibitem [{\citenamefont {DiVincenzo}(2000)}]{P.:2000aa}%
  \BibitemOpen
  \bibfield  {author} {\bibinfo {author} {\bibfnamefont {D.~P.}\ \bibnamefont
  {DiVincenzo}},\ }\bibfield  {booktitle} {\emph {\bibinfo {booktitle}
  {Fortschritte der Physik}},\ }\href {\doibase
  10.1002/1521-3978(200009)48:9/11<771::AID-PROP771>3.0.CO;2-E} {\bibfield
  {journal} {\bibinfo  {journal} {Fortschritte der Physik}\ }\textbf {\bibinfo
  {volume} {48}},\ \bibinfo {pages} {771} (\bibinfo {year} {2000})}\BibitemShut
  {NoStop}%
\bibitem [{\citenamefont {Song}\ \emph {et~al.}(2017)\citenamefont {Song},
  \citenamefont {Xu}, \citenamefont {Liu}, \citenamefont {Yang}, \citenamefont
  {Zheng}, \citenamefont {Deng}, \citenamefont {Xie}, \citenamefont {Huang},
  \citenamefont {Guo}, \citenamefont {Zhang}, \citenamefont {Zhang},
  \citenamefont {Xu}, \citenamefont {Zheng}, \citenamefont {Zhu}, \citenamefont
  {Wang}, \citenamefont {Chen}, \citenamefont {Lu}, \citenamefont {Han},\ and\
  \citenamefont {Pan}}]{Song:2017aa}%
  \BibitemOpen
  \bibfield  {author} {\bibinfo {author} {\bibfnamefont {C.}~\bibnamefont
  {Song}}, \bibinfo {author} {\bibfnamefont {K.}~\bibnamefont {Xu}}, \bibinfo
  {author} {\bibfnamefont {W.}~\bibnamefont {Liu}}, \bibinfo {author}
  {\bibfnamefont {C.-p.}\ \bibnamefont {Yang}}, \bibinfo {author}
  {\bibfnamefont {S.-B.}\ \bibnamefont {Zheng}}, \bibinfo {author}
  {\bibfnamefont {H.}~\bibnamefont {Deng}}, \bibinfo {author} {\bibfnamefont
  {Q.}~\bibnamefont {Xie}}, \bibinfo {author} {\bibfnamefont {K.}~\bibnamefont
  {Huang}}, \bibinfo {author} {\bibfnamefont {Q.}~\bibnamefont {Guo}}, \bibinfo
  {author} {\bibfnamefont {L.}~\bibnamefont {Zhang}}, \bibinfo {author}
  {\bibfnamefont {P.}~\bibnamefont {Zhang}}, \bibinfo {author} {\bibfnamefont
  {D.}~\bibnamefont {Xu}}, \bibinfo {author} {\bibfnamefont {D.}~\bibnamefont
  {Zheng}}, \bibinfo {author} {\bibfnamefont {X.}~\bibnamefont {Zhu}}, \bibinfo
  {author} {\bibfnamefont {H.}~\bibnamefont {Wang}}, \bibinfo {author}
  {\bibfnamefont {Y.~A.}\ \bibnamefont {Chen}}, \bibinfo {author}
  {\bibfnamefont {C.~Y.}\ \bibnamefont {Lu}}, \bibinfo {author} {\bibfnamefont
  {S.}~\bibnamefont {Han}}, \ and\ \bibinfo {author} {\bibfnamefont {J.-W.}\
  \bibnamefont {Pan}},\ }\href {\doibase 10.1103/PhysRevLett.119.180511}
  {\bibfield  {journal} {\bibinfo  {journal} {Physical Review Letters}\
  }\textbf {\bibinfo {volume} {119}},\ \bibinfo {pages} {180511} (\bibinfo
  {year} {2017})}\BibitemShut {NoStop}%
\bibitem [{\citenamefont {Sch{\"a}fer}\ \emph {et~al.}(2018)\citenamefont
  {Sch{\"a}fer}, \citenamefont {Ballance}, \citenamefont {Thirumalai},
  \citenamefont {Stephenson}, \citenamefont {Ballance}, \citenamefont
  {Steane},\ and\ \citenamefont {Lucas}}]{Schafer:2018aa}%
  \BibitemOpen
  \bibfield  {author} {\bibinfo {author} {\bibfnamefont {V.~M.}\ \bibnamefont
  {Sch{\"a}fer}}, \bibinfo {author} {\bibfnamefont {C.~J.}\ \bibnamefont
  {Ballance}}, \bibinfo {author} {\bibfnamefont {K.}~\bibnamefont
  {Thirumalai}}, \bibinfo {author} {\bibfnamefont {L.~J.}\ \bibnamefont
  {Stephenson}}, \bibinfo {author} {\bibfnamefont {T.~G.}\ \bibnamefont
  {Ballance}}, \bibinfo {author} {\bibfnamefont {A.~M.}\ \bibnamefont
  {Steane}}, \ and\ \bibinfo {author} {\bibfnamefont {D.~M.}\ \bibnamefont
  {Lucas}},\ }\href {http://dx.doi.org/10.1038/nature25737} {\bibfield
  {journal} {\bibinfo  {journal} {Nature}\ }\textbf {\bibinfo {volume} {555}},\
  \bibinfo {pages} {75} (\bibinfo {year} {2018})}\BibitemShut {NoStop}%
\bibitem [{\citenamefont {Press}\ \emph {et~al.}(2008)\citenamefont {Press},
  \citenamefont {Ladd}, \citenamefont {Zhang},\ and\ \citenamefont
  {Yamamoto}}]{Press:2008aa}%
  \BibitemOpen
  \bibfield  {author} {\bibinfo {author} {\bibfnamefont {D.}~\bibnamefont
  {Press}}, \bibinfo {author} {\bibfnamefont {T.~D.}\ \bibnamefont {Ladd}},
  \bibinfo {author} {\bibfnamefont {B.}~\bibnamefont {Zhang}}, \ and\ \bibinfo
  {author} {\bibfnamefont {Y.}~\bibnamefont {Yamamoto}},\ }\href
  {http://dx.doi.org/10.1038/nature07530} {\bibfield  {journal} {\bibinfo
  {journal} {Nature}\ }\textbf {\bibinfo {volume} {456}},\ \bibinfo {pages}
  {218} (\bibinfo {year} {2008})}\BibitemShut {NoStop}%
\bibitem [{\citenamefont {Delteil}\ \emph {et~al.}(2017)\citenamefont
  {Delteil}, \citenamefont {Gao}, \citenamefont {Sun},\ and\ \citenamefont
  {Imamo{\u g}lu}}]{Delteil:2017aa}%
  \BibitemOpen
  \bibfield  {author} {\bibinfo {author} {\bibfnamefont {A.}~\bibnamefont
  {Delteil}}, \bibinfo {author} {\bibfnamefont {W.-b.}\ \bibnamefont {Gao}},
  \bibinfo {author} {\bibfnamefont {Z.}~\bibnamefont {Sun}}, \ and\ \bibinfo
  {author} {\bibfnamefont {A.}~\bibnamefont {Imamo{\u g}lu}},\ }\enquote
  {\bibinfo {title} {Entanglement generation based on quantum dot spins},}\ in\
  \href {\doibase 10.1007/978-3-319-56378-7{\_}12} {\emph {\bibinfo {booktitle}
  {Quantum Dots for Quantum Information Technologies}}},\ \bibinfo {editor}
  {edited by\ \bibinfo {editor} {\bibfnamefont {P.}~\bibnamefont {Michler}}}\
  (\bibinfo  {publisher} {Springer International Publishing},\ \bibinfo
  {address} {Cham},\ \bibinfo {year} {2017})\ pp.\ \bibinfo {pages}
  {379--407}\BibitemShut {NoStop}%
\bibitem [{\citenamefont {Caspani}\ \emph {et~al.}(2017)\citenamefont
  {Caspani}, \citenamefont {Xiong}, \citenamefont {Eggleton}, \citenamefont
  {Bajoni}, \citenamefont {Liscidini}, \citenamefont {Galli}, \citenamefont
  {Morandotti},\ and\ \citenamefont {Moss}}]{Caspani:2017aa}%
  \BibitemOpen
  \bibfield  {author} {\bibinfo {author} {\bibfnamefont {L.}~\bibnamefont
  {Caspani}}, \bibinfo {author} {\bibfnamefont {C.}~\bibnamefont {Xiong}},
  \bibinfo {author} {\bibfnamefont {B.~J.}\ \bibnamefont {Eggleton}}, \bibinfo
  {author} {\bibfnamefont {D.}~\bibnamefont {Bajoni}}, \bibinfo {author}
  {\bibfnamefont {M.}~\bibnamefont {Liscidini}}, \bibinfo {author}
  {\bibfnamefont {M.}~\bibnamefont {Galli}}, \bibinfo {author} {\bibfnamefont
  {R.}~\bibnamefont {Morandotti}}, \ and\ \bibinfo {author} {\bibfnamefont
  {D.~J.}\ \bibnamefont {Moss}},\ }\href
  {http://dx.doi.org/10.1038/lsa.2017.100} {\bibfield  {journal} {\bibinfo
  {journal} {Light: Science \&Amp; Applications}\ }\textbf {\bibinfo {volume}
  {6}},\ \bibinfo {pages} {e17100} (\bibinfo {year} {2017})}\BibitemShut
  {NoStop}%
\bibitem [{\citenamefont {Bar-Gill}\ \emph {et~al.}(2013)\citenamefont
  {Bar-Gill}, \citenamefont {Pham}, \citenamefont {Jarmola}, \citenamefont
  {Budker},\ and\ \citenamefont {Walsworth}}]{Bar-Gill:2013aa}%
  \BibitemOpen
  \bibfield  {author} {\bibinfo {author} {\bibfnamefont {N.}~\bibnamefont
  {Bar-Gill}}, \bibinfo {author} {\bibfnamefont {L.~M.}\ \bibnamefont {Pham}},
  \bibinfo {author} {\bibfnamefont {A.}~\bibnamefont {Jarmola}}, \bibinfo
  {author} {\bibfnamefont {D.}~\bibnamefont {Budker}}, \ and\ \bibinfo {author}
  {\bibfnamefont {R.~L.}\ \bibnamefont {Walsworth}},\ }\href
  {http://dx.doi.org/10.1038/ncomms2771} {\bibfield  {journal} {\bibinfo
  {journal} {Nature Communications}\ }\textbf {\bibinfo {volume} {4}},\
  \bibinfo {pages} {1743} (\bibinfo {year} {2013})}\BibitemShut {NoStop}%
\bibitem [{\citenamefont {de~Lange}\ \emph {et~al.}(2010)\citenamefont
  {de~Lange}, \citenamefont {Wang}, \citenamefont {Rist{\`e}}, \citenamefont
  {Dobrovitski},\ and\ \citenamefont {Hanson}}]{Lange:2010aa}%
  \BibitemOpen
  \bibfield  {author} {\bibinfo {author} {\bibfnamefont {G.}~\bibnamefont
  {de~Lange}}, \bibinfo {author} {\bibfnamefont {Z.~H.}\ \bibnamefont {Wang}},
  \bibinfo {author} {\bibfnamefont {D.}~\bibnamefont {Rist{\`e}}}, \bibinfo
  {author} {\bibfnamefont {V.~V.}\ \bibnamefont {Dobrovitski}}, \ and\ \bibinfo
  {author} {\bibfnamefont {R.}~\bibnamefont {Hanson}},\ }\href
  {http://science.sciencemag.org/content/330/6000/60.abstract} {\bibfield
  {journal} {\bibinfo  {journal} {Science}\ }\textbf {\bibinfo {volume}
  {330}},\ \bibinfo {pages} {60} (\bibinfo {year} {2010})}\BibitemShut
  {NoStop}%
\bibitem [{\citenamefont {Neumann}\ \emph
  {et~al.}(2010{\natexlab{a}})\citenamefont {Neumann}, \citenamefont {Beck},
  \citenamefont {Steiner}, \citenamefont {Rempp}, \citenamefont {Fedder},
  \citenamefont {Hemmer}, \citenamefont {Wrachtrup},\ and\ \citenamefont
  {Jelezko}}]{Neumann:2010aa}%
  \BibitemOpen
  \bibfield  {author} {\bibinfo {author} {\bibfnamefont {P.}~\bibnamefont
  {Neumann}}, \bibinfo {author} {\bibfnamefont {J.}~\bibnamefont {Beck}},
  \bibinfo {author} {\bibfnamefont {M.}~\bibnamefont {Steiner}}, \bibinfo
  {author} {\bibfnamefont {F.}~\bibnamefont {Rempp}}, \bibinfo {author}
  {\bibfnamefont {H.}~\bibnamefont {Fedder}}, \bibinfo {author} {\bibfnamefont
  {P.~R.}\ \bibnamefont {Hemmer}}, \bibinfo {author} {\bibfnamefont
  {J.}~\bibnamefont {Wrachtrup}}, \ and\ \bibinfo {author} {\bibfnamefont
  {F.}~\bibnamefont {Jelezko}},\ }\href
  {http://science.sciencemag.org/content/329/5991/542.abstract} {\bibfield
  {journal} {\bibinfo  {journal} {Science}\ }\textbf {\bibinfo {volume}
  {329}},\ \bibinfo {pages} {542} (\bibinfo {year}
  {2010}{\natexlab{a}})}\BibitemShut {NoStop}%
\bibitem [{\citenamefont {Jelezko}\ \emph {et~al.}(2004)\citenamefont
  {Jelezko}, \citenamefont {Gaebel}, \citenamefont {Popa}, \citenamefont
  {Gruber},\ and\ \citenamefont {Wrachtrup}}]{Jelezko:2004aa}%
  \BibitemOpen
  \bibfield  {author} {\bibinfo {author} {\bibfnamefont {F.}~\bibnamefont
  {Jelezko}}, \bibinfo {author} {\bibfnamefont {T.}~\bibnamefont {Gaebel}},
  \bibinfo {author} {\bibfnamefont {I.}~\bibnamefont {Popa}}, \bibinfo {author}
  {\bibfnamefont {A.}~\bibnamefont {Gruber}}, \ and\ \bibinfo {author}
  {\bibfnamefont {J.}~\bibnamefont {Wrachtrup}},\ }\href {\doibase
  10.1103/PhysRevLett.92.076401} {\bibfield  {journal} {\bibinfo  {journal}
  {Physical Review Letters}\ }\textbf {\bibinfo {volume} {92}},\ \bibinfo
  {pages} {076401} (\bibinfo {year} {2004})}\BibitemShut {NoStop}%
\bibitem [{\citenamefont {Gruber}\ \emph {et~al.}(1997)\citenamefont {Gruber},
  \citenamefont {Dr{\"a}benstedt}, \citenamefont {Tietz}, \citenamefont
  {Fleury}, \citenamefont {Wrachtrup},\ and\ \citenamefont
  {Borczyskowski}}]{Gruber:1997aa}%
  \BibitemOpen
  \bibfield  {author} {\bibinfo {author} {\bibfnamefont {A.}~\bibnamefont
  {Gruber}}, \bibinfo {author} {\bibfnamefont {A.}~\bibnamefont
  {Dr{\"a}benstedt}}, \bibinfo {author} {\bibfnamefont {C.}~\bibnamefont
  {Tietz}}, \bibinfo {author} {\bibfnamefont {L.}~\bibnamefont {Fleury}},
  \bibinfo {author} {\bibfnamefont {J.}~\bibnamefont {Wrachtrup}}, \ and\
  \bibinfo {author} {\bibfnamefont {C.~v.}\ \bibnamefont {Borczyskowski}},\
  }\href {http://science.sciencemag.org/content/276/5321/2012.abstract}
  {\bibfield  {journal} {\bibinfo  {journal} {Science}\ }\textbf {\bibinfo
  {volume} {276}},\ \bibinfo {pages} {2012} (\bibinfo {year}
  {1997})}\BibitemShut {NoStop}%
\bibitem [{\citenamefont {Hanson}\ \emph {et~al.}(2006)\citenamefont {Hanson},
  \citenamefont {Mendoza}, \citenamefont {Epstein},\ and\ \citenamefont
  {Awschalom}}]{Hanson:2006aa}%
  \BibitemOpen
  \bibfield  {author} {\bibinfo {author} {\bibfnamefont {R.}~\bibnamefont
  {Hanson}}, \bibinfo {author} {\bibfnamefont {F.~M.}\ \bibnamefont {Mendoza}},
  \bibinfo {author} {\bibfnamefont {R.~J.}\ \bibnamefont {Epstein}}, \ and\
  \bibinfo {author} {\bibfnamefont {D.~D.}\ \bibnamefont {Awschalom}},\ }\href
  {\doibase 10.1103/PhysRevLett.97.087601} {\bibfield  {journal} {\bibinfo
  {journal} {Physical Review Letters}\ }\textbf {\bibinfo {volume} {97}},\
  \bibinfo {pages} {087601} (\bibinfo {year} {2006})}\BibitemShut {NoStop}%
\bibitem [{\citenamefont {Hensen}\ \emph {et~al.}(2015)\citenamefont {Hensen},
  \citenamefont {Bernien}, \citenamefont {Dr{\'e}au}, \citenamefont {Reiserer},
  \citenamefont {Kalb}, \citenamefont {Blok}, \citenamefont {Ruitenberg},
  \citenamefont {Vermeulen}, \citenamefont {Schouten}, \citenamefont
  {Abell{\'a}n}, \citenamefont {Amaya}, \citenamefont {Pruneri}, \citenamefont
  {Mitchell}, \citenamefont {Markham}, \citenamefont {Twitchen}, \citenamefont
  {Elkouss}, \citenamefont {Wehner}, \citenamefont {Taminiau},\ and\
  \citenamefont {Hanson}}]{Hensen:2015aa}%
  \BibitemOpen
  \bibfield  {author} {\bibinfo {author} {\bibfnamefont {B.}~\bibnamefont
  {Hensen}}, \bibinfo {author} {\bibfnamefont {H.}~\bibnamefont {Bernien}},
  \bibinfo {author} {\bibfnamefont {A.~E.}\ \bibnamefont {Dr{\'e}au}}, \bibinfo
  {author} {\bibfnamefont {A.}~\bibnamefont {Reiserer}}, \bibinfo {author}
  {\bibfnamefont {N.}~\bibnamefont {Kalb}}, \bibinfo {author} {\bibfnamefont
  {M.~S.}\ \bibnamefont {Blok}}, \bibinfo {author} {\bibfnamefont
  {J.}~\bibnamefont {Ruitenberg}}, \bibinfo {author} {\bibfnamefont {R.~F.~L.}\
  \bibnamefont {Vermeulen}}, \bibinfo {author} {\bibfnamefont {R.~N.}\
  \bibnamefont {Schouten}}, \bibinfo {author} {\bibfnamefont {C.}~\bibnamefont
  {Abell{\'a}n}}, \bibinfo {author} {\bibfnamefont {W.}~\bibnamefont {Amaya}},
  \bibinfo {author} {\bibfnamefont {V.}~\bibnamefont {Pruneri}}, \bibinfo
  {author} {\bibfnamefont {M.~W.}\ \bibnamefont {Mitchell}}, \bibinfo {author}
  {\bibfnamefont {M.}~\bibnamefont {Markham}}, \bibinfo {author} {\bibfnamefont
  {D.~J.}\ \bibnamefont {Twitchen}}, \bibinfo {author} {\bibfnamefont
  {D.}~\bibnamefont {Elkouss}}, \bibinfo {author} {\bibfnamefont
  {S.}~\bibnamefont {Wehner}}, \bibinfo {author} {\bibfnamefont {T.~H.}\
  \bibnamefont {Taminiau}}, \ and\ \bibinfo {author} {\bibfnamefont
  {R.}~\bibnamefont {Hanson}},\ }\href {http://dx.doi.org/10.1038/nature15759}
  {\bibfield  {journal} {\bibinfo  {journal} {Nature}\ }\textbf {\bibinfo
  {volume} {526}},\ \bibinfo {pages} {682} (\bibinfo {year}
  {2015})}\BibitemShut {NoStop}%
\bibitem [{\citenamefont {Humphreys}\ \emph {et~al.}(2018)\citenamefont
  {Humphreys}, \citenamefont {Kalb}, \citenamefont {Morits}, \citenamefont
  {Schouten}, \citenamefont {Vermeulen}, \citenamefont {Twitchen},
  \citenamefont {Markham},\ and\ \citenamefont {Hanson}}]{Humphreys2017}%
  \BibitemOpen
  \bibfield  {author} {\bibinfo {author} {\bibfnamefont {P.~C.}\ \bibnamefont
  {Humphreys}}, \bibinfo {author} {\bibfnamefont {N.}~\bibnamefont {Kalb}},
  \bibinfo {author} {\bibfnamefont {J.~P.~J.}\ \bibnamefont {Morits}}, \bibinfo
  {author} {\bibfnamefont {R.~N.}\ \bibnamefont {Schouten}}, \bibinfo {author}
  {\bibfnamefont {R.~F.~L.}\ \bibnamefont {Vermeulen}}, \bibinfo {author}
  {\bibfnamefont {D.~J.}\ \bibnamefont {Twitchen}}, \bibinfo {author}
  {\bibfnamefont {M.}~\bibnamefont {Markham}}, \ and\ \bibinfo {author}
  {\bibfnamefont {R.}~\bibnamefont {Hanson}},\ }\href {\doibase
  10.1038/s41586-018-0200-5} {\bibfield  {journal} {\bibinfo  {journal}
  {Nature}\ }\textbf {\bibinfo {volume} {558}},\ \bibinfo {pages} {268}
  (\bibinfo {year} {2018})}\BibitemShut {NoStop}%
\bibitem [{\citenamefont {Mamin}\ \emph {et~al.}(2013)\citenamefont {Mamin},
  \citenamefont {Kim}, \citenamefont {Sherwood}, \citenamefont {Rettner},
  \citenamefont {Ohno}, \citenamefont {Awschalom},\ and\ \citenamefont
  {Rugar}}]{Mamin:2013aa}%
  \BibitemOpen
  \bibfield  {author} {\bibinfo {author} {\bibfnamefont {H.~J.}\ \bibnamefont
  {Mamin}}, \bibinfo {author} {\bibfnamefont {M.}~\bibnamefont {Kim}}, \bibinfo
  {author} {\bibfnamefont {M.~H.}\ \bibnamefont {Sherwood}}, \bibinfo {author}
  {\bibfnamefont {C.~T.}\ \bibnamefont {Rettner}}, \bibinfo {author}
  {\bibfnamefont {K.}~\bibnamefont {Ohno}}, \bibinfo {author} {\bibfnamefont
  {D.~D.}\ \bibnamefont {Awschalom}}, \ and\ \bibinfo {author} {\bibfnamefont
  {D.}~\bibnamefont {Rugar}},\ }\href
  {http://science.sciencemag.org/content/339/6119/557.abstract} {\bibfield
  {journal} {\bibinfo  {journal} {Science}\ }\textbf {\bibinfo {volume}
  {339}},\ \bibinfo {pages} {557} (\bibinfo {year} {2013})}\BibitemShut
  {NoStop}%
\bibitem [{\citenamefont {Maurer}\ \emph {et~al.}(2012)\citenamefont {Maurer},
  \citenamefont {Kucsko}, \citenamefont {Latta}, \citenamefont {Jiang},
  \citenamefont {Yao}, \citenamefont {Bennett}, \citenamefont {Pastawski},
  \citenamefont {Hunger}, \citenamefont {Chisholm}, \citenamefont {Markham},
  \citenamefont {Twitchen}, \citenamefont {Cirac},\ and\ \citenamefont
  {Lukin}}]{Maurer:2012aa}%
  \BibitemOpen
  \bibfield  {author} {\bibinfo {author} {\bibfnamefont {P.~C.}\ \bibnamefont
  {Maurer}}, \bibinfo {author} {\bibfnamefont {G.}~\bibnamefont {Kucsko}},
  \bibinfo {author} {\bibfnamefont {C.}~\bibnamefont {Latta}}, \bibinfo
  {author} {\bibfnamefont {L.}~\bibnamefont {Jiang}}, \bibinfo {author}
  {\bibfnamefont {N.~Y.}\ \bibnamefont {Yao}}, \bibinfo {author} {\bibfnamefont
  {S.~D.}\ \bibnamefont {Bennett}}, \bibinfo {author} {\bibfnamefont
  {F.}~\bibnamefont {Pastawski}}, \bibinfo {author} {\bibfnamefont
  {D.}~\bibnamefont {Hunger}}, \bibinfo {author} {\bibfnamefont
  {N.}~\bibnamefont {Chisholm}}, \bibinfo {author} {\bibfnamefont
  {M.}~\bibnamefont {Markham}}, \bibinfo {author} {\bibfnamefont {D.~J.}\
  \bibnamefont {Twitchen}}, \bibinfo {author} {\bibfnamefont {J.~I.}\
  \bibnamefont {Cirac}}, \ and\ \bibinfo {author} {\bibfnamefont {M.~D.}\
  \bibnamefont {Lukin}},\ }\href
  {http://science.sciencemag.org/content/336/6086/1283.abstract} {\bibfield
  {journal} {\bibinfo  {journal} {Science}\ }\textbf {\bibinfo {volume}
  {336}},\ \bibinfo {pages} {1283} (\bibinfo {year} {2012})}\BibitemShut
  {NoStop}%
\bibitem [{\citenamefont {Yao}\ \emph {et~al.}(2011)\citenamefont {Yao},
  \citenamefont {Jiang}, \citenamefont {Gorshkov}, \citenamefont {Gong},
  \citenamefont {Zhai}, \citenamefont {Duan},\ and\ \citenamefont
  {Lukin}}]{Yao:2011aa}%
  \BibitemOpen
  \bibfield  {author} {\bibinfo {author} {\bibfnamefont {N.~Y.}\ \bibnamefont
  {Yao}}, \bibinfo {author} {\bibfnamefont {L.}~\bibnamefont {Jiang}}, \bibinfo
  {author} {\bibfnamefont {A.~V.}\ \bibnamefont {Gorshkov}}, \bibinfo {author}
  {\bibfnamefont {Z.~X.}\ \bibnamefont {Gong}}, \bibinfo {author}
  {\bibfnamefont {A.}~\bibnamefont {Zhai}}, \bibinfo {author} {\bibfnamefont
  {L.~M.}\ \bibnamefont {Duan}}, \ and\ \bibinfo {author} {\bibfnamefont
  {M.~D.}\ \bibnamefont {Lukin}},\ }\href {\doibase
  10.1103/PhysRevLett.106.040505} {\bibfield  {journal} {\bibinfo  {journal}
  {Physical Review Letters}\ }\textbf {\bibinfo {volume} {106}},\ \bibinfo
  {pages} {040505} (\bibinfo {year} {2011})}\BibitemShut {NoStop}%
\bibitem [{\citenamefont {Cai}\ \emph {et~al.}(2013)\citenamefont {Cai},
  \citenamefont {Retzker}, \citenamefont {Jelezko},\ and\ \citenamefont
  {Plenio}}]{Cai:2013aa}%
  \BibitemOpen
  \bibfield  {author} {\bibinfo {author} {\bibfnamefont {J.}~\bibnamefont
  {Cai}}, \bibinfo {author} {\bibfnamefont {A.}~\bibnamefont {Retzker}},
  \bibinfo {author} {\bibfnamefont {F.}~\bibnamefont {Jelezko}}, \ and\
  \bibinfo {author} {\bibfnamefont {M.~B.}\ \bibnamefont {Plenio}},\ }\href
  {http://dx.doi.org/10.1038/nphys2519} {\bibfield  {journal} {\bibinfo
  {journal} {Nature Physics}\ }\textbf {\bibinfo {volume} {9}},\ \bibinfo
  {pages} {168} (\bibinfo {year} {2013})}\BibitemShut {NoStop}%
\bibitem [{\citenamefont {Wang}\ \emph {et~al.}(2015)\citenamefont {Wang},
  \citenamefont {Dolde}, \citenamefont {Biamonte}, \citenamefont {Babbush},
  \citenamefont {Bergholm}, \citenamefont {Yang}, \citenamefont {Jakobi},
  \citenamefont {Neumann}, \citenamefont {Aspuru-Guzik}, \citenamefont
  {Whitfield},\ and\ \citenamefont {Wrachtrup}}]{Wang:2015aa}%
  \BibitemOpen
  \bibfield  {author} {\bibinfo {author} {\bibfnamefont {Y.}~\bibnamefont
  {Wang}}, \bibinfo {author} {\bibfnamefont {F.}~\bibnamefont {Dolde}},
  \bibinfo {author} {\bibfnamefont {J.}~\bibnamefont {Biamonte}}, \bibinfo
  {author} {\bibfnamefont {R.}~\bibnamefont {Babbush}}, \bibinfo {author}
  {\bibfnamefont {V.}~\bibnamefont {Bergholm}}, \bibinfo {author}
  {\bibfnamefont {S.}~\bibnamefont {Yang}}, \bibinfo {author} {\bibfnamefont
  {I.}~\bibnamefont {Jakobi}}, \bibinfo {author} {\bibfnamefont
  {P.}~\bibnamefont {Neumann}}, \bibinfo {author} {\bibfnamefont
  {A.}~\bibnamefont {Aspuru-Guzik}}, \bibinfo {author} {\bibfnamefont {J.~D.}\
  \bibnamefont {Whitfield}}, \ and\ \bibinfo {author} {\bibfnamefont
  {J.}~\bibnamefont {Wrachtrup}},\ }\bibfield  {booktitle} {\emph {\bibinfo
  {booktitle} {ACS Nano}},\ }\href {\doibase 10.1021/acsnano.5b01651}
  {\bibfield  {journal} {\bibinfo  {journal} {ACS Nano}\ }\textbf {\bibinfo
  {volume} {9}},\ \bibinfo {pages} {7769} (\bibinfo {year} {2015})}\BibitemShut
  {NoStop}%
\bibitem [{\citenamefont {Rong}\ \emph {et~al.}(2015)\citenamefont {Rong},
  \citenamefont {Geng}, \citenamefont {Shi}, \citenamefont {Liu}, \citenamefont
  {Xu}, \citenamefont {Ma}, \citenamefont {Kong}, \citenamefont {Jiang},
  \citenamefont {Wu},\ and\ \citenamefont {Du}}]{Rong:2015aa}%
  \BibitemOpen
  \bibfield  {author} {\bibinfo {author} {\bibfnamefont {X.}~\bibnamefont
  {Rong}}, \bibinfo {author} {\bibfnamefont {J.}~\bibnamefont {Geng}}, \bibinfo
  {author} {\bibfnamefont {F.}~\bibnamefont {Shi}}, \bibinfo {author}
  {\bibfnamefont {Y.}~\bibnamefont {Liu}}, \bibinfo {author} {\bibfnamefont
  {K.}~\bibnamefont {Xu}}, \bibinfo {author} {\bibfnamefont {W.}~\bibnamefont
  {Ma}}, \bibinfo {author} {\bibfnamefont {F.}~\bibnamefont {Kong}}, \bibinfo
  {author} {\bibfnamefont {Z.}~\bibnamefont {Jiang}}, \bibinfo {author}
  {\bibfnamefont {Y.}~\bibnamefont {Wu}}, \ and\ \bibinfo {author}
  {\bibfnamefont {J.}~\bibnamefont {Du}},\ }\href
  {http://dx.doi.org/10.1038/ncomms9748} {\bibfield  {journal} {\bibinfo
  {journal} {Nature Communications}\ }\textbf {\bibinfo {volume} {6}},\
  \bibinfo {pages} {8748} (\bibinfo {year} {2015})}\BibitemShut {NoStop}%
\bibitem [{\citenamefont {Dolde}\ \emph {et~al.}(2013)\citenamefont {Dolde},
  \citenamefont {Jakobi}, \citenamefont {Naydenov}, \citenamefont {Zhao},
  \citenamefont {Pezzagna}, \citenamefont {Trautmann}, \citenamefont {Meijer},
  \citenamefont {Neumann}, \citenamefont {Jelezko},\ and\ \citenamefont
  {Wrachtrup}}]{Dolde:2013aa}%
  \BibitemOpen
  \bibfield  {author} {\bibinfo {author} {\bibfnamefont {F.}~\bibnamefont
  {Dolde}}, \bibinfo {author} {\bibfnamefont {I.}~\bibnamefont {Jakobi}},
  \bibinfo {author} {\bibfnamefont {B.}~\bibnamefont {Naydenov}}, \bibinfo
  {author} {\bibfnamefont {N.}~\bibnamefont {Zhao}}, \bibinfo {author}
  {\bibfnamefont {S.}~\bibnamefont {Pezzagna}}, \bibinfo {author}
  {\bibfnamefont {C.}~\bibnamefont {Trautmann}}, \bibinfo {author}
  {\bibfnamefont {J.}~\bibnamefont {Meijer}}, \bibinfo {author} {\bibfnamefont
  {P.}~\bibnamefont {Neumann}}, \bibinfo {author} {\bibfnamefont
  {F.}~\bibnamefont {Jelezko}}, \ and\ \bibinfo {author} {\bibfnamefont
  {J.}~\bibnamefont {Wrachtrup}},\ }\href {http://dx.doi.org/10.1038/nphys2545}
  {\bibfield  {journal} {\bibinfo  {journal} {Nature Physics}\ }\textbf
  {\bibinfo {volume} {9}},\ \bibinfo {pages} {139} (\bibinfo {year}
  {2013})}\BibitemShut {NoStop}%
\bibitem [{\citenamefont {Dolde}\ \emph {et~al.}(2014)\citenamefont {Dolde},
  \citenamefont {Bergholm}, \citenamefont {Wang}, \citenamefont {Jakobi},
  \citenamefont {Naydenov}, \citenamefont {Pezzagna}, \citenamefont {Meijer},
  \citenamefont {Jelezko}, \citenamefont {Neumann}, \citenamefont
  {Schulte-Herbr{\"u}ggen}, \citenamefont {Biamonte},\ and\ \citenamefont
  {Wrachtrup}}]{Dolde:2014aa}%
  \BibitemOpen
  \bibfield  {author} {\bibinfo {author} {\bibfnamefont {F.}~\bibnamefont
  {Dolde}}, \bibinfo {author} {\bibfnamefont {V.}~\bibnamefont {Bergholm}},
  \bibinfo {author} {\bibfnamefont {Y.}~\bibnamefont {Wang}}, \bibinfo {author}
  {\bibfnamefont {I.}~\bibnamefont {Jakobi}}, \bibinfo {author} {\bibfnamefont
  {B.}~\bibnamefont {Naydenov}}, \bibinfo {author} {\bibfnamefont
  {S.}~\bibnamefont {Pezzagna}}, \bibinfo {author} {\bibfnamefont
  {J.}~\bibnamefont {Meijer}}, \bibinfo {author} {\bibfnamefont
  {F.}~\bibnamefont {Jelezko}}, \bibinfo {author} {\bibfnamefont
  {P.}~\bibnamefont {Neumann}}, \bibinfo {author} {\bibfnamefont
  {T.}~\bibnamefont {Schulte-Herbr{\"u}ggen}}, \bibinfo {author} {\bibfnamefont
  {J.}~\bibnamefont {Biamonte}}, \ and\ \bibinfo {author} {\bibfnamefont
  {J.}~\bibnamefont {Wrachtrup}},\ }\href
  {http://dx.doi.org/10.1038/ncomms4371} {\bibfield  {journal} {\bibinfo
  {journal} {Nature Communications}\ }\textbf {\bibinfo {volume} {5}},\
  \bibinfo {pages} {3371} (\bibinfo {year} {2014})}\BibitemShut {NoStop}%
\bibitem [{\citenamefont {Yao}\ \emph {et~al.}(2012)\citenamefont {Yao},
  \citenamefont {Jiang}, \citenamefont {Gorshkov}, \citenamefont {Maurer},
  \citenamefont {Giedke}, \citenamefont {Cirac},\ and\ \citenamefont
  {Lukin}}]{NYao2012scalable}%
  \BibitemOpen
  \bibfield  {author} {\bibinfo {author} {\bibfnamefont {N.}~\bibnamefont
  {Yao}}, \bibinfo {author} {\bibfnamefont {L.}~\bibnamefont {Jiang}}, \bibinfo
  {author} {\bibfnamefont {A.}~\bibnamefont {Gorshkov}}, \bibinfo {author}
  {\bibfnamefont {P.}~\bibnamefont {Maurer}}, \bibinfo {author} {\bibfnamefont
  {G.}~\bibnamefont {Giedke}}, \bibinfo {author} {\bibfnamefont
  {J.}~\bibnamefont {Cirac}}, \ and\ \bibinfo {author} {\bibfnamefont
  {M.}~\bibnamefont {Lukin}},\ }\href {\doibase 10.1038/ncomms1788} {\bibfield
  {journal} {\bibinfo  {journal} {Nature Communications}\ }\textbf {\bibinfo
  {volume} {3}},\ \bibinfo {pages} {800} (\bibinfo {year} {2012})}\BibitemShut
  {NoStop}%
\bibitem [{\citenamefont {Greentree}(2016)}]{Greentree:2016aa}%
  \BibitemOpen
  \bibfield  {author} {\bibinfo {author} {\bibfnamefont {A.~D.}\ \bibnamefont
  {Greentree}},\ }\href {http://stacks.iop.org/1367-2630/18/i=2/a=021002}
  {\bibfield  {journal} {\bibinfo  {journal} {New Journal of Physics}\ }\textbf
  {\bibinfo {volume} {18}},\ \bibinfo {pages} {021002} (\bibinfo {year}
  {2016})}\BibitemShut {NoStop}%
\bibitem [{\citenamefont {Schr{\"o}der}\ \emph {et~al.}(2017)\citenamefont
  {Schr{\"o}der}, \citenamefont {Walsh}, \citenamefont {Zheng}, \citenamefont
  {Mouradian}, \citenamefont {Li}, \citenamefont {Malladi}, \citenamefont
  {Bakhru}, \citenamefont {Lu}, \citenamefont {Stein}, \citenamefont {Heuck},\
  and\ \citenamefont {Englund}}]{Schroder:2017aa}%
  \BibitemOpen
  \bibfield  {author} {\bibinfo {author} {\bibfnamefont {T.}~\bibnamefont
  {Schr{\"o}der}}, \bibinfo {author} {\bibfnamefont {M.}~\bibnamefont {Walsh}},
  \bibinfo {author} {\bibfnamefont {J.}~\bibnamefont {Zheng}}, \bibinfo
  {author} {\bibfnamefont {S.}~\bibnamefont {Mouradian}}, \bibinfo {author}
  {\bibfnamefont {L.}~\bibnamefont {Li}}, \bibinfo {author} {\bibfnamefont
  {G.}~\bibnamefont {Malladi}}, \bibinfo {author} {\bibfnamefont
  {H.}~\bibnamefont {Bakhru}}, \bibinfo {author} {\bibfnamefont
  {M.}~\bibnamefont {Lu}}, \bibinfo {author} {\bibfnamefont {A.}~\bibnamefont
  {Stein}}, \bibinfo {author} {\bibfnamefont {M.}~\bibnamefont {Heuck}}, \ and\
  \bibinfo {author} {\bibfnamefont {D.}~\bibnamefont {Englund}},\ }\bibfield
  {booktitle} {\emph {\bibinfo {booktitle} {Optical Materials Express}},\
  }\href {\doibase 10.1364/OME.7.001514} {\bibfield  {journal} {\bibinfo
  {journal} {Optical Materials Express}\ }\textbf {\bibinfo {volume} {7}},\
  \bibinfo {pages} {1514} (\bibinfo {year} {2017})}\BibitemShut {NoStop}%
\bibitem [{\citenamefont {Bersin}\ \emph {et~al.}(2019)\citenamefont {Bersin},
  \citenamefont {Walsh}, \citenamefont {Mouradian}, \citenamefont {Trusheim},
  \citenamefont {Schr{\"o}der},\ and\ \citenamefont {Englund}}]{Bersin2019}%
  \BibitemOpen
  \bibfield  {author} {\bibinfo {author} {\bibfnamefont {E.}~\bibnamefont
  {Bersin}}, \bibinfo {author} {\bibfnamefont {M.}~\bibnamefont {Walsh}},
  \bibinfo {author} {\bibfnamefont {S.~L.}\ \bibnamefont {Mouradian}}, \bibinfo
  {author} {\bibfnamefont {M.~E.}\ \bibnamefont {Trusheim}}, \bibinfo {author}
  {\bibfnamefont {T.}~\bibnamefont {Schr{\"o}der}}, \ and\ \bibinfo {author}
  {\bibfnamefont {D.}~\bibnamefont {Englund}},\ }\href {\doibase
  10.1038/s41534-019-0154-y} {\bibfield  {journal} {\bibinfo  {journal} {npj
  Quantum Information}\ }\textbf {\bibinfo {volume} {5}},\ \bibinfo {pages}
  {38} (\bibinfo {year} {2019})}\BibitemShut {NoStop}%
\bibitem [{\citenamefont {Abobeih}\ \emph {et~al.}(2018)\citenamefont
  {Abobeih}, \citenamefont {Cramer}, \citenamefont {Bakker}, \citenamefont
  {Kalb}, \citenamefont {Markham}, \citenamefont {Twitchen},\ and\
  \citenamefont {Taminiau}}]{Abobeih2018}%
  \BibitemOpen
  \bibfield  {author} {\bibinfo {author} {\bibfnamefont {M.~H.}\ \bibnamefont
  {Abobeih}}, \bibinfo {author} {\bibfnamefont {J.}~\bibnamefont {Cramer}},
  \bibinfo {author} {\bibfnamefont {M.~A.}\ \bibnamefont {Bakker}}, \bibinfo
  {author} {\bibfnamefont {N.}~\bibnamefont {Kalb}}, \bibinfo {author}
  {\bibfnamefont {M.}~\bibnamefont {Markham}}, \bibinfo {author} {\bibfnamefont
  {D.~J.}\ \bibnamefont {Twitchen}}, \ and\ \bibinfo {author} {\bibfnamefont
  {T.~H.}\ \bibnamefont {Taminiau}},\ }\href@noop {} {\bibfield  {journal}
  {\bibinfo  {journal} {Nature Communications}\ }\textbf {\bibinfo {volume}
  {9}},\ \bibinfo {pages} {2552} (\bibinfo {year} {2018})}\BibitemShut
  {NoStop}%
\bibitem [{\citenamefont {Jarmola}\ \emph {et~al.}(2012)\citenamefont
  {Jarmola}, \citenamefont {Acosta}, \citenamefont {Jensen}, \citenamefont
  {Chemerisov},\ and\ \citenamefont {Budker}}]{Jarmola2012T1}%
  \BibitemOpen
  \bibfield  {author} {\bibinfo {author} {\bibfnamefont {A.}~\bibnamefont
  {Jarmola}}, \bibinfo {author} {\bibfnamefont {V.~M.}\ \bibnamefont {Acosta}},
  \bibinfo {author} {\bibfnamefont {K.}~\bibnamefont {Jensen}}, \bibinfo
  {author} {\bibfnamefont {S.}~\bibnamefont {Chemerisov}}, \ and\ \bibinfo
  {author} {\bibfnamefont {D.}~\bibnamefont {Budker}},\ }\href@noop {}
  {\bibfield  {journal} {\bibinfo  {journal} {Phys. Rev. Lett.}\ }\textbf
  {\bibinfo {volume} {108}},\ \bibinfo {pages} {197601} (\bibinfo {year}
  {2012})}\BibitemShut {NoStop}%
\bibitem [{\citenamefont {Ma}\ \emph {et~al.}(2017)\citenamefont {Ma},
  \citenamefont {Hoang}, \citenamefont {Gong}, \citenamefont {Li},\ and\
  \citenamefont {Yin}}]{Ma2017}%
  \BibitemOpen
  \bibfield  {author} {\bibinfo {author} {\bibfnamefont {Y.}~\bibnamefont
  {Ma}}, \bibinfo {author} {\bibfnamefont {T.~M.}\ \bibnamefont {Hoang}},
  \bibinfo {author} {\bibfnamefont {M.}~\bibnamefont {Gong}}, \bibinfo {author}
  {\bibfnamefont {T.}~\bibnamefont {Li}}, \ and\ \bibinfo {author}
  {\bibfnamefont {Z.-q.}\ \bibnamefont {Yin}},\ }\href {\doibase
  10.1103/PhysRevA.96.023827} {\bibfield  {journal} {\bibinfo  {journal} {Phys.
  Rev. A}\ }\textbf {\bibinfo {volume} {96}},\ \bibinfo {pages} {023827}
  (\bibinfo {year} {2017})}\BibitemShut {NoStop}%
\bibitem [{\citenamefont {Reynhardt}\ \emph {et~al.}(1998)\citenamefont
  {Reynhardt}, \citenamefont {High},\ and\ \citenamefont {van
  Wyk}}]{Reynhardt:1998aa}%
  \BibitemOpen
  \bibfield  {author} {\bibinfo {author} {\bibfnamefont {E.~C.}\ \bibnamefont
  {Reynhardt}}, \bibinfo {author} {\bibfnamefont {G.~L.}\ \bibnamefont {High}},
  \ and\ \bibinfo {author} {\bibfnamefont {J.~A.}\ \bibnamefont {van Wyk}},\
  }\href {\doibase 10.1063/1.477511} {\bibfield  {journal} {\bibinfo  {journal}
  {The Journal of Chemical Physics}\ }\textbf {\bibinfo {volume} {109}},\
  \bibinfo {pages} {8471} (\bibinfo {year} {1998})},\ \Eprint
  {http://arxiv.org/abs/https://doi.org/10.1063/1.477511}
  {https://doi.org/10.1063/1.477511} \BibitemShut {NoStop}%
\bibitem [{\citenamefont {Rondin}\ \emph {et~al.}(2014)\citenamefont {Rondin},
  \citenamefont {Tetienne}, \citenamefont {Hingant}, \citenamefont {Roch},
  \citenamefont {Maletinsky},\ and\ \citenamefont {Jacques}}]{Rondin:2014aa}%
  \BibitemOpen
  \bibfield  {author} {\bibinfo {author} {\bibfnamefont {L.}~\bibnamefont
  {Rondin}}, \bibinfo {author} {\bibfnamefont {J.-P.}\ \bibnamefont
  {Tetienne}}, \bibinfo {author} {\bibfnamefont {T.}~\bibnamefont {Hingant}},
  \bibinfo {author} {\bibfnamefont {J.-F.}\ \bibnamefont {Roch}}, \bibinfo
  {author} {\bibfnamefont {P.}~\bibnamefont {Maletinsky}}, \ and\ \bibinfo
  {author} {\bibfnamefont {V.}~\bibnamefont {Jacques}},\ }\href
  {http://stacks.iop.org/0034-4885/77/i=5/a=056503} {\bibfield  {journal}
  {\bibinfo  {journal} {Reports on Progress in Physics}\ }\textbf {\bibinfo
  {volume} {77}},\ \bibinfo {pages} {056503} (\bibinfo {year}
  {2014})}\BibitemShut {NoStop}%
\bibitem [{\citenamefont {Doherty}\ \emph {et~al.}(2012)\citenamefont
  {Doherty}, \citenamefont {Dolde}, \citenamefont {Fedder}, \citenamefont
  {Jelezko}, \citenamefont {Wrachtrup}, \citenamefont {Manson},\ and\
  \citenamefont {Hollenberg}}]{Doherty:2012aa}%
  \BibitemOpen
  \bibfield  {author} {\bibinfo {author} {\bibfnamefont {M.~W.}\ \bibnamefont
  {Doherty}}, \bibinfo {author} {\bibfnamefont {F.}~\bibnamefont {Dolde}},
  \bibinfo {author} {\bibfnamefont {H.}~\bibnamefont {Fedder}}, \bibinfo
  {author} {\bibfnamefont {F.}~\bibnamefont {Jelezko}}, \bibinfo {author}
  {\bibfnamefont {J.}~\bibnamefont {Wrachtrup}}, \bibinfo {author}
  {\bibfnamefont {N.~B.}\ \bibnamefont {Manson}}, \ and\ \bibinfo {author}
  {\bibfnamefont {L.~C.~L.}\ \bibnamefont {Hollenberg}},\ }\href {\doibase
  10.1103/PhysRevB.85.205203} {\bibfield  {journal} {\bibinfo  {journal} {Phys.
  Rev. B}\ }\textbf {\bibinfo {volume} {85}},\ \bibinfo {pages} {205203}
  (\bibinfo {year} {2012})}\BibitemShut {NoStop}%
\bibitem [{\citenamefont {Tamarat}\ \emph {et~al.}(2008)\citenamefont
  {Tamarat}, \citenamefont {Manson}, \citenamefont {Harrison}, \citenamefont
  {McMurtrie}, \citenamefont {Nizovtsev}, \citenamefont {Beausoleil},
  \citenamefont {Neumann}, \citenamefont {Gaebel}, \citenamefont {Jelezko},
  \citenamefont {Hemmer},\ and\ \citenamefont {Wrachtrup}}]{Tamarat2008}%
  \BibitemOpen
  \bibfield  {author} {\bibinfo {author} {\bibfnamefont {P.}~\bibnamefont
  {Tamarat}}, \bibinfo {author} {\bibfnamefont {N.~B.}\ \bibnamefont {Manson}},
  \bibinfo {author} {\bibfnamefont {J.~P.}\ \bibnamefont {Harrison}}, \bibinfo
  {author} {\bibfnamefont {R.~L.}\ \bibnamefont {McMurtrie}}, \bibinfo {author}
  {\bibfnamefont {C.}~\bibnamefont {Nizovtsev}, \bibfnamefont {A.and~Santori}},
  \bibinfo {author} {\bibfnamefont {R.}~\bibnamefont {Beausoleil}}, \bibinfo
  {author} {\bibfnamefont {P.}~\bibnamefont {Neumann}}, \bibinfo {author}
  {\bibfnamefont {T.}~\bibnamefont {Gaebel}}, \bibinfo {author} {\bibfnamefont
  {F.}~\bibnamefont {Jelezko}}, \bibinfo {author} {\bibfnamefont
  {P.}~\bibnamefont {Hemmer}}, \ and\ \bibinfo {author} {\bibfnamefont
  {J.}~\bibnamefont {Wrachtrup}},\ }\href@noop {} {\bibfield  {journal}
  {\bibinfo  {journal} {New Journal of Physics}\ }\textbf {\bibinfo {volume}
  {10}},\ \bibinfo {pages} {1367} (\bibinfo {year} {2008})}\BibitemShut
  {NoStop}%
\bibitem [{\citenamefont {Lee}\ \emph {et~al.}(2016)\citenamefont {Lee},
  \citenamefont {Lee}, \citenamefont {Ovartchaiyapong}, \citenamefont
  {Minguzzi}, \citenamefont {Maze},\ and\ \citenamefont
  {Bleszynski~Jayich}}]{Lee2016strain}%
  \BibitemOpen
  \bibfield  {author} {\bibinfo {author} {\bibfnamefont {K.~W.}\ \bibnamefont
  {Lee}}, \bibinfo {author} {\bibfnamefont {D.}~\bibnamefont {Lee}}, \bibinfo
  {author} {\bibfnamefont {P.}~\bibnamefont {Ovartchaiyapong}}, \bibinfo
  {author} {\bibfnamefont {J.}~\bibnamefont {Minguzzi}}, \bibinfo {author}
  {\bibfnamefont {J.~R.}\ \bibnamefont {Maze}}, \ and\ \bibinfo {author}
  {\bibfnamefont {A.~C.}\ \bibnamefont {Bleszynski~Jayich}},\ }\href {\doibase
  10.1103/PhysRevApplied.6.034005} {\bibfield  {journal} {\bibinfo  {journal}
  {Phys. Rev. Applied}\ }\textbf {\bibinfo {volume} {6}},\ \bibinfo {pages}
  {034005} (\bibinfo {year} {2016})}\BibitemShut {NoStop}%
\bibitem [{\citenamefont {Doherty}\ and\ \citenamefont {{\em et
  al}.}(2014)}]{doherty2014pressure}%
  \BibitemOpen
  \bibfield  {author} {\bibinfo {author} {\bibfnamefont {M.~W.}\ \bibnamefont
  {Doherty}}\ and\ \bibinfo {author} {\bibnamefont {{\em et al}.}},\
  }\href@noop {} {\bibfield  {journal} {\bibinfo  {journal} {Phys. Rev. Lett.}\
  }\textbf {\bibinfo {volume} {112}},\ \bibinfo {pages} {047601} (\bibinfo
  {year} {2014})}\BibitemShut {NoStop}%
\bibitem [{\citenamefont {Manson}\ \emph {et~al.}(2006)\citenamefont {Manson},
  \citenamefont {Harrison},\ and\ \citenamefont {Sellars}}]{Manson:2006aa}%
  \BibitemOpen
  \bibfield  {author} {\bibinfo {author} {\bibfnamefont {N.~B.}\ \bibnamefont
  {Manson}}, \bibinfo {author} {\bibfnamefont {J.~P.}\ \bibnamefont
  {Harrison}}, \ and\ \bibinfo {author} {\bibfnamefont {M.~J.}\ \bibnamefont
  {Sellars}},\ }\href {https://link.aps.org/doi/10.1103/PhysRevB.74.104303}
  {\bibfield  {journal} {\bibinfo  {journal} {Physical Review B}\ }\textbf
  {\bibinfo {volume} {74}},\ \bibinfo {pages} {104303} (\bibinfo {year}
  {2006})}\BibitemShut {NoStop}%
\bibitem [{\citenamefont {Tamarat}\ \emph {et~al.}(2006)\citenamefont
  {Tamarat}, \citenamefont {Gaebel}, \citenamefont {Rabeau}, \citenamefont
  {Khan}, \citenamefont {Greentree}, \citenamefont {Wilson}, \citenamefont
  {Hollenberg}, \citenamefont {Prawer}, \citenamefont {Hemmer}, \citenamefont
  {Jelezko},\ and\ \citenamefont {Wrachtrup}}]{Tamarat2006linewidth}%
  \BibitemOpen
  \bibfield  {author} {\bibinfo {author} {\bibfnamefont {P.}~\bibnamefont
  {Tamarat}}, \bibinfo {author} {\bibfnamefont {T.}~\bibnamefont {Gaebel}},
  \bibinfo {author} {\bibfnamefont {J.~R.}\ \bibnamefont {Rabeau}}, \bibinfo
  {author} {\bibfnamefont {M.}~\bibnamefont {Khan}}, \bibinfo {author}
  {\bibfnamefont {A.~D.}\ \bibnamefont {Greentree}}, \bibinfo {author}
  {\bibfnamefont {H.}~\bibnamefont {Wilson}}, \bibinfo {author} {\bibfnamefont
  {L.~C.~L.}\ \bibnamefont {Hollenberg}}, \bibinfo {author} {\bibfnamefont
  {S.}~\bibnamefont {Prawer}}, \bibinfo {author} {\bibfnamefont
  {P.}~\bibnamefont {Hemmer}}, \bibinfo {author} {\bibfnamefont
  {F.}~\bibnamefont {Jelezko}}, \ and\ \bibinfo {author} {\bibfnamefont
  {J.}~\bibnamefont {Wrachtrup}},\ }\href@noop {} {\bibfield  {journal}
  {\bibinfo  {journal} {Phys. Rev. Lett.}\ }\textbf {\bibinfo {volume} {97}},\
  \bibinfo {pages} {083002} (\bibinfo {year} {2006})}\BibitemShut {NoStop}%
\bibitem [{\citenamefont {Babinec}\ \emph {et~al.}(2011)\citenamefont
  {Babinec}, \citenamefont {Choy}, \citenamefont {Smith}, \citenamefont
  {Khan},\ and\ \citenamefont {Lon{\v c}ar}}]{Babinec:2011aa}%
  \BibitemOpen
  \bibfield  {author} {\bibinfo {author} {\bibfnamefont {T.~M.}\ \bibnamefont
  {Babinec}}, \bibinfo {author} {\bibfnamefont {J.~T.}\ \bibnamefont {Choy}},
  \bibinfo {author} {\bibfnamefont {K.~J.~M.}\ \bibnamefont {Smith}}, \bibinfo
  {author} {\bibfnamefont {M.}~\bibnamefont {Khan}}, \ and\ \bibinfo {author}
  {\bibfnamefont {M.}~\bibnamefont {Lon{\v c}ar}},\ }\bibfield  {booktitle}
  {\emph {\bibinfo {booktitle} {Journal of Vacuum Science \& Technology B,
  Nanotechnology and Microelectronics: Materials, Processing, Measurement, and
  Phenomena}},\ }\href {\doibase 10.1116/1.3520638} {\bibfield  {journal}
  {\bibinfo  {journal} {Journal of Vacuum Science \& Technology B,
  Nanotechnology and Microelectronics: Materials, Processing, Measurement, and
  Phenomena}\ }\textbf {\bibinfo {volume} {29}},\ \bibinfo {pages} {010601}
  (\bibinfo {year} {2011})}\BibitemShut {NoStop}%
\bibitem [{\citenamefont {Neumann}\ \emph
  {et~al.}(2010{\natexlab{b}})\citenamefont {Neumann}, \citenamefont {Kolesov},
  \citenamefont {Naydenov}, \citenamefont {Beck}, \citenamefont {Rempp},
  \citenamefont {Steiner}, \citenamefont {Jacques}, \citenamefont
  {Balasubramanian}, \citenamefont {Markham}, \citenamefont {Twitchen},
  \citenamefont {Pezzagna}, \citenamefont {Meijer}, \citenamefont {Twamley},
  \citenamefont {Jelezko},\ and\ \citenamefont {Wrachtrup}}]{Neumann:2010ab}%
  \BibitemOpen
  \bibfield  {author} {\bibinfo {author} {\bibfnamefont {P.}~\bibnamefont
  {Neumann}}, \bibinfo {author} {\bibfnamefont {R.}~\bibnamefont {Kolesov}},
  \bibinfo {author} {\bibfnamefont {B.}~\bibnamefont {Naydenov}}, \bibinfo
  {author} {\bibfnamefont {J.}~\bibnamefont {Beck}}, \bibinfo {author}
  {\bibfnamefont {F.}~\bibnamefont {Rempp}}, \bibinfo {author} {\bibfnamefont
  {M.}~\bibnamefont {Steiner}}, \bibinfo {author} {\bibfnamefont
  {V.}~\bibnamefont {Jacques}}, \bibinfo {author} {\bibfnamefont
  {G.}~\bibnamefont {Balasubramanian}}, \bibinfo {author} {\bibfnamefont
  {M.~L.}\ \bibnamefont {Markham}}, \bibinfo {author} {\bibfnamefont {D.~J.}\
  \bibnamefont {Twitchen}}, \bibinfo {author} {\bibfnamefont {S.}~\bibnamefont
  {Pezzagna}}, \bibinfo {author} {\bibfnamefont {J.}~\bibnamefont {Meijer}},
  \bibinfo {author} {\bibfnamefont {J.}~\bibnamefont {Twamley}}, \bibinfo
  {author} {\bibfnamefont {F.}~\bibnamefont {Jelezko}}, \ and\ \bibinfo
  {author} {\bibfnamefont {J.}~\bibnamefont {Wrachtrup}},\ }\href
  {http://dx.doi.org/10.1038/nphys1536} {\bibfield  {journal} {\bibinfo
  {journal} {Nature Physics}\ }\textbf {\bibinfo {volume} {6}},\ \bibinfo
  {pages} {249} (\bibinfo {year} {2010}{\natexlab{b}})}\BibitemShut {NoStop}%
\bibitem [{\citenamefont {Nielsen}\ and\ \citenamefont
  {Chuang}(2010)}]{Nielsen:2010aa}%
  \BibitemOpen
  \bibfield  {author} {\bibinfo {author} {\bibfnamefont {M.~A.}\ \bibnamefont
  {Nielsen}}\ and\ \bibinfo {author} {\bibfnamefont {I.~L.}\ \bibnamefont
  {Chuang}},\ }\href {\doibase DOI: 10.1017/CBO9780511976667} {\emph {\bibinfo
  {title} {Quantum Computation and Quantum Information: 10th Anniversary
  Edition}}}\ (\bibinfo  {publisher} {Cambridge University Press},\ \bibinfo
  {address} {Cambridge},\ \bibinfo {year} {2010})\BibitemShut {NoStop}%
\bibitem [{\citenamefont {Li}\ \emph {et~al.}(2018)\citenamefont {Li},
  \citenamefont {Ma}, \citenamefont {Han}, \citenamefont {Chen}, \citenamefont
  {Xu}, \citenamefont {Cai}, \citenamefont {Wang}, \citenamefont {Song},
  \citenamefont {Xue}, \citenamefont {Yin},\ and\ \citenamefont
  {Sun}}]{2018arXiv180603886L}%
  \BibitemOpen
  \bibfield  {author} {\bibinfo {author} {\bibfnamefont {X.}~\bibnamefont
  {Li}}, \bibinfo {author} {\bibfnamefont {Y.}~\bibnamefont {Ma}}, \bibinfo
  {author} {\bibfnamefont {J.}~\bibnamefont {Han}}, \bibinfo {author}
  {\bibfnamefont {T.}~\bibnamefont {Chen}}, \bibinfo {author} {\bibfnamefont
  {Y.}~\bibnamefont {Xu}}, \bibinfo {author} {\bibfnamefont {W.}~\bibnamefont
  {Cai}}, \bibinfo {author} {\bibfnamefont {H.}~\bibnamefont {Wang}}, \bibinfo
  {author} {\bibfnamefont {Y.}~\bibnamefont {Song}}, \bibinfo {author}
  {\bibfnamefont {Z.-Y.}\ \bibnamefont {Xue}}, \bibinfo {author} {\bibfnamefont
  {Z.-q.}\ \bibnamefont {Yin}}, \ and\ \bibinfo {author} {\bibfnamefont
  {L.}~\bibnamefont {Sun}},\ }\href {\doibase 10.1103/PhysRevApplied.10.054009}
  {\bibfield  {journal} {\bibinfo  {journal} {Phys. Rev. Applied}\ }\textbf
  {\bibinfo {volume} {10}},\ \bibinfo {pages} {054009} (\bibinfo {year}
  {2018})}\BibitemShut {NoStop}%
\bibitem [{\citenamefont {Yu}\ \emph {et~al.}(2013)\citenamefont {Yu},
  \citenamefont {Duan},\ and\ \citenamefont {Ying}}]{Yu:2013aa}%
  \BibitemOpen
  \bibfield  {author} {\bibinfo {author} {\bibfnamefont {N.}~\bibnamefont
  {Yu}}, \bibinfo {author} {\bibfnamefont {R.}~\bibnamefont {Duan}}, \ and\
  \bibinfo {author} {\bibfnamefont {M.}~\bibnamefont {Ying}},\ }\href {\doibase
  10.1103/PhysRevA.88.010304} {\bibfield  {journal} {\bibinfo  {journal}
  {Physical Review A}\ }\textbf {\bibinfo {volume} {88}},\ \bibinfo {pages}
  {010304} (\bibinfo {year} {2013})}\BibitemShut {NoStop}%
\bibitem [{\citenamefont {Machnes}\ \emph {et~al.}(2011)\citenamefont
  {Machnes}, \citenamefont {Sander}, \citenamefont {Glaser}, \citenamefont
  {de~Fouqui{\`e}res}, \citenamefont {Gruslys}, \citenamefont {Schirmer},\ and\
  \citenamefont {Schulte-Herbr{\"u}ggen}}]{Machnes:2011aa}%
  \BibitemOpen
  \bibfield  {author} {\bibinfo {author} {\bibfnamefont {S.}~\bibnamefont
  {Machnes}}, \bibinfo {author} {\bibfnamefont {U.}~\bibnamefont {Sander}},
  \bibinfo {author} {\bibfnamefont {S.~J.}\ \bibnamefont {Glaser}}, \bibinfo
  {author} {\bibfnamefont {P.}~\bibnamefont {de~Fouqui{\`e}res}}, \bibinfo
  {author} {\bibfnamefont {A.}~\bibnamefont {Gruslys}}, \bibinfo {author}
  {\bibfnamefont {S.}~\bibnamefont {Schirmer}}, \ and\ \bibinfo {author}
  {\bibfnamefont {T.}~\bibnamefont {Schulte-Herbr{\"u}ggen}},\ }\href {\doibase
  10.1103/PhysRevA.84.022305} {\bibfield  {journal} {\bibinfo  {journal}
  {Physical Review A}\ }\textbf {\bibinfo {volume} {84}},\ \bibinfo {pages}
  {022305} (\bibinfo {year} {2011})}\BibitemShut {NoStop}%
\bibitem [{\citenamefont {Waldherr}\ \emph {et~al.}(2014)\citenamefont
  {Waldherr}, \citenamefont {Wang}, \citenamefont {Zaiser}, \citenamefont
  {Jamali}, \citenamefont {Schulte-Herbr{\"u}ggen}, \citenamefont {Abe},
  \citenamefont {Ohshima}, \citenamefont {Isoya}, \citenamefont {Du},
  \citenamefont {Neumann},\ and\ \citenamefont {Wrachtrup}}]{Waldherr:2014aa}%
  \BibitemOpen
  \bibfield  {author} {\bibinfo {author} {\bibfnamefont {G.}~\bibnamefont
  {Waldherr}}, \bibinfo {author} {\bibfnamefont {Y.}~\bibnamefont {Wang}},
  \bibinfo {author} {\bibfnamefont {S.}~\bibnamefont {Zaiser}}, \bibinfo
  {author} {\bibfnamefont {M.}~\bibnamefont {Jamali}}, \bibinfo {author}
  {\bibfnamefont {T.}~\bibnamefont {Schulte-Herbr{\"u}ggen}}, \bibinfo {author}
  {\bibfnamefont {H.}~\bibnamefont {Abe}}, \bibinfo {author} {\bibfnamefont
  {T.}~\bibnamefont {Ohshima}}, \bibinfo {author} {\bibfnamefont
  {J.}~\bibnamefont {Isoya}}, \bibinfo {author} {\bibfnamefont {J.~F.}\
  \bibnamefont {Du}}, \bibinfo {author} {\bibfnamefont {P.}~\bibnamefont
  {Neumann}}, \ and\ \bibinfo {author} {\bibfnamefont {J.}~\bibnamefont
  {Wrachtrup}},\ }\href {http://dx.doi.org/10.1038/nature12919} {\bibfield
  {journal} {\bibinfo  {journal} {Nature}\ }\textbf {\bibinfo {volume} {506}},\
  \bibinfo {pages} {204} (\bibinfo {year} {2014})}\BibitemShut {NoStop}%
\bibitem [{\citenamefont {Khaneja}\ \emph {et~al.}(2005)\citenamefont
  {Khaneja}, \citenamefont {Reiss}, \citenamefont {Kehlet}, \citenamefont
  {Schulte-Herbr{\"u}ggen},\ and\ \citenamefont {Glaser}}]{Khaneja:2005aa}%
  \BibitemOpen
  \bibfield  {author} {\bibinfo {author} {\bibfnamefont {N.}~\bibnamefont
  {Khaneja}}, \bibinfo {author} {\bibfnamefont {T.}~\bibnamefont {Reiss}},
  \bibinfo {author} {\bibfnamefont {C.}~\bibnamefont {Kehlet}}, \bibinfo
  {author} {\bibfnamefont {T.}~\bibnamefont {Schulte-Herbr{\"u}ggen}}, \ and\
  \bibinfo {author} {\bibfnamefont {S.~J.}\ \bibnamefont {Glaser}},\ }\href
  {\doibase https://doi.org/10.1016/j.jmr.2004.11.004} {\bibfield  {journal}
  {\bibinfo  {journal} {Journal of Magnetic Resonance}\ }\textbf {\bibinfo
  {volume} {172}},\ \bibinfo {pages} {296} (\bibinfo {year}
  {2005})}\BibitemShut {NoStop}%
\bibitem [{\citenamefont {Nielsen}(2002)}]{Nielsen:2002aa}%
  \BibitemOpen
  \bibfield  {author} {\bibinfo {author} {\bibfnamefont {M.~A.}\ \bibnamefont
  {Nielsen}},\ }\href {\doibase https://doi.org/10.1016/S0375-9601(02)01272-0}
  {\bibfield  {journal} {\bibinfo  {journal} {Physics Letters A}\ }\textbf
  {\bibinfo {volume} {303}},\ \bibinfo {pages} {249} (\bibinfo {year}
  {2002})}\BibitemShut {NoStop}%
\bibitem [{\citenamefont {Bowdrey}\ \emph {et~al.}(2002)\citenamefont
  {Bowdrey}, \citenamefont {Oi}, \citenamefont {Short}, \citenamefont
  {Banaszek},\ and\ \citenamefont {Jones}}]{Bowdrey:2002aa}%
  \BibitemOpen
  \bibfield  {author} {\bibinfo {author} {\bibfnamefont {M.~D.}\ \bibnamefont
  {Bowdrey}}, \bibinfo {author} {\bibfnamefont {D.~K.~L.}\ \bibnamefont {Oi}},
  \bibinfo {author} {\bibfnamefont {A.~J.}\ \bibnamefont {Short}}, \bibinfo
  {author} {\bibfnamefont {K.}~\bibnamefont {Banaszek}}, \ and\ \bibinfo
  {author} {\bibfnamefont {J.~A.}\ \bibnamefont {Jones}},\ }\href {\doibase
  https://doi.org/10.1016/S0375-9601(02)00069-5} {\bibfield  {journal}
  {\bibinfo  {journal} {Physics Letters A}\ }\textbf {\bibinfo {volume}
  {294}},\ \bibinfo {pages} {258} (\bibinfo {year} {2002})}\BibitemShut
  {NoStop}%
\bibitem [{\citenamefont {Loss}\ and\ \citenamefont
  {DiVincenzo}(1998)}]{Loss:1998aa}%
  \BibitemOpen
  \bibfield  {author} {\bibinfo {author} {\bibfnamefont {D.}~\bibnamefont
  {Loss}}\ and\ \bibinfo {author} {\bibfnamefont {D.~P.}\ \bibnamefont
  {DiVincenzo}},\ }\href {https://link.aps.org/doi/10.1103/PhysRevA.57.120}
  {\bibfield  {journal} {\bibinfo  {journal} {Physical Review A}\ }\textbf
  {\bibinfo {volume} {57}},\ \bibinfo {pages} {120} (\bibinfo {year}
  {1998})}\BibitemShut {NoStop}%
\end{thebibliography}

%

\end{document}